\documentclass[aps,twocolumn,superscriptaddress,nofootinbib,a4paper,longbibliography]{revtex4-2}
\pdfoutput=1

\usepackage{times}
\usepackage{epsfig}
\usepackage{amsfonts}
\usepackage{amsmath}
\usepackage{amsthm}
\usepackage{amssymb}					
\usepackage{amsthm}
\usepackage{dsfont}
\usepackage{bm}
\usepackage{mathtools}
\usepackage{color}
\usepackage{multirow}
\usepackage[normalem]{ulem}
\newcommand{\stkout}[1]{\ifmmode\text{\sout{\ensuremath{#1}}}\else\sout{#1}\fi}
\usepackage{latexsym}
\usepackage{mathrsfs}
\usepackage{natbib}
\usepackage{verbatim}
\usepackage[T1]{fontenc}
\usepackage{float}
\usepackage{graphicx}
\usepackage{xcolor}

\newtheorem{theorem}{Theorem}
\newtheorem{proposition}[theorem]{Proposition}

\theoremstyle{definition}
\newtheorem{definition}[theorem]{Definition}

\DeclareMathOperator{\Tr}{tr}

\newcommand{\id}{\mathds{1}}

\usepackage[colorlinks=true,linkcolor=blue,citecolor=magenta,urlcolor=blue]{hyperref}

\begin{document}


\title{Bounding and Simulating Contextual Correlations in Quantum Theory}


\author{Armin Tavakoli}
\affiliation{Institute for Quantum Optics and Quantum Information -- IQOQI Vienna, Austrian Academy of Sciences, Boltzmanngasse 3, 1090 Vienna, Austria}
\affiliation{D\'epartement de Physique Appliqu\'ee, Universit\'e de Gen\`eve, CH-1211 Gen\`eve, Switzerland}
\author{Emmanuel Zambrini Cruzeiro}
\affiliation{Laboratoire d'Information Quantique, CP 225, Universit\'e libre de Bruxelles (ULB), Av. F. D. Roosevelt 50, 1050 Bruxelles, Belgium}

\author{Roope Uola}
\affiliation{D\'epartement de Physique Appliqu\'ee, Universit\'e de Gen\`eve, CH-1211 Gen\`eve, Switzerland}

\author{Alastair A.\ Abbott}
\affiliation{D\'epartement de Physique Appliqu\'ee, Universit\'e de Gen\`eve, CH-1211 Gen\`eve, Switzerland}
\affiliation{Univ.\ Grenoble Alpes, Inria, 38000 Grenoble, France}

\begin{abstract}
We introduce a hierarchy of semidefinite relaxations of the set of quantum correlations in generalised contextuality scenarios. This constitutes a simple and versatile tool for bounding the magnitude of quantum contextuality. 
To illustrate its utility, we use it to determine the maximal quantum violation of several noncontextuality inequalities whose maximum violations were previously unknown.
We then go further and use it to prove that certain preparation-contextual correlations cannot be explained with pure states, thereby showing that mixed states are an indispensable resource for contextuality. 
In the second part of the paper, we turn our attention to the simulation of preparation-contextual correlations in general operational theories.
We introduce the information cost of simulating preparation contextuality, which quantifies the additional, otherwise forbidden, information required to simulate contextual correlations in either classical or quantum models.
In both cases, we show that the simulation cost can be efficiently bounded using a variant of our hierarchy of semidefinite relaxations, and we calculate it exactly in the simplest contextuality scenario of parity-oblivious multiplexing.
\end{abstract}


\maketitle


\section{Introduction}

The contextuality of quantum theory is a fundamental sign of its nonclassicality that has been investigated for several decades. 
While contextuality was originally established as a property specific to the formalism of quantum theory~\cite{BellContext, KS}, it has, in more recent times, been further generalised as a property of nonclassical probability distributions that can arise in operational theories~\cite{Spekkens}. 
This operational notion of contextuality is applicable to a broad range of physical scenarios and has been shown to  be linked to a variety of foundational and applied topics in quantum theory (see, e.g., Refs.~\cite{Spekkens2,Pusey,Leifer,Lostaglio,ArminRoope,Lostaglio2,Anwer,Saha2019,Kunjwal}).

The principle of noncontextuality holds that operationally equivalent physical procedures must correspond to identical descriptions in any underlying ontological model~\cite{Spekkens}. 
This assumption imposes constraints on the correlations that can be obtained in prepare-and-measure scenarios involving operationally equivalent preparations and measurements.
In such scenarios, which we term ``contextuality scenarios'', the correlations obtainable by noncontextual models can be characterised in terms of linear programming~\cite{NCpolytope}. 
In contrast, quantum models that nonetheless respect the operational equivalences may produce ``contextual correlations'' unobtainable by any such noncontextual model~\cite{Spekkens}.
This leads to a conceptually natural question: how can we determine if, for a given contextuality scenario, a given set of contextual correlations is compatible with quantum theory?
This question is crucial for understanding the extent of nonclassicality manifested in quantum theory, and hence also for the development of quantum information protocols powered by quantum contextuality.
While an explicit quantum model is sufficient to prove compatibility with quantum theory, proving the converse---that no such model exists---is more challenging. 

Here, we provide an answer to the question by introducing a hierarchy of semidefinite relaxations of the set of quantum correlations arising in contextuality scenarios involving arbitrary operational equivalences between preparations and measurements. 
This constitutes a sequence of increasingly precise necessary conditions that contextual correlations must satisfy in order to admit of a quantum model. 
Thus, if a given contextual probability distribution fails one of the tests, it is incompatible with any quantum model satisfying the specified operational equivalences. 
We exemplify their practical utility by determining the maximal quantum violations of several different noncontextuality inequalities (for noisy state discrimination~\cite{Schmid}, for three-dimensional parity-oblivious multiplexing~\cite{Ambainis}, for the communication task experimentally investigated in Ref.~\cite{Hameedi}, and for the polytope inequalities obtained in Ref.~\cite{NCpolytope}). 
Then, we apply our method to solve a foundational problem in quantum contextuality: we present a correlation inequality satisfied by all quantum models based on pure states and show that it can be violated by quantum strategies exploiting mixed states. 
Thus, we prove that mixed states are an indispensable resource for strong forms of quantum contextuality.

Equipped with the ability to bound the magnitude of quantum contextuality, we ask what additional resources are required to simulate preparation contextual correlations with classical or quantum models.
We identify this resource as the preparation of states deviating from the required operational equivalences, and quantify this deviation in terms of the information extractable about the operational equivalences via measurement.
This allows us to interpret preparation contextuality scenarios, and experiments aiming to simulate their results, as particular types of informationally restricted correlation experiments~\cite{Info1, Info2}.  
For both classical and quantum models, we show that the simulation cost can be lower bounded using variants of our hierarchy of semidefinite relaxations. We apply these concepts to the simplest preparation contextuality scenario~\cite{POM}, where we explicitly derive both the classical and quantum simulation costs of contextuality.

\section{Contextuality}

Consider a prepare-and-measure experiment in which Alice receives an input  $x\in [n_X] \coloneqq \{1,\ldots,n_X\}$, prepares a system using the preparation $P_x$ and sends it to Bob. 
Bob receives an input $y\in[n_Y]$, performs a measurement $M_y$ and obtains an outcome $b\in[n_B]$; this event is called the \emph{measurement effect} and is denoted $[b|M_y]$. 
When the experiment is repeated many times it gives rise to the conditional probability distribution $p(b|x,y)\coloneqq p(b|P_x,M_y)$.
	
An ontological model provides a realist explanation of the observed correlations $p(b|x,y)$~\cite{Spekkens}. In an ontological model, the preparation is associated to an ontic variable $\lambda$ subject to some distribution (i.e., an epistemic state) $p(\lambda|x)$ and the measurement is represented by a probabilistic response function depending on the ontic state, $p(b|y,\lambda)$.%
\footnote{\label{fn:CL}These distributions must, respectively, be linear in preparations $x$ (since the epistemic state of a mixture of preparations is the mixture of the epistemic states of the respective preparations) and measurement effects $[b|M_y]$ (since a measurement effect arising from a mixture of measurements or a post-processing of outcomes must be represented by the corresponding mixture and post-processing of response functions).} 
The observed correlations are then written
\begin{equation}
p(b|x,y)=\sum_{\lambda} p(\lambda|x)p(b|y,\lambda).
\end{equation}
Notice that every probability distribution admits of an ontological model.

\subsection{Operational equivalences}

The notion of noncontextuality becomes relevant when certain operational procedures (either preparations or measurements) are \emph{operationally equivalent}~\cite{Spekkens}.
Two preparations $P$ and $P'$ are said to be operationally equivalent, denoted $P\simeq P'$, if no measurement\footnote{Here, the quantifier is over all possible measurements (that, e.g., Bob could perform), not only the fixed set $\{M_y\}_y$ he uses in the prepare-and-measure experiment at hand.} can distinguish them, i.e.,
\begin{equation}
	\forall\, [b|M]: \quad p(b|P,M) = p(b|P',M).
\end{equation}
Similarly, two measurement effects $[b|M]$ and $[b'|M']$ are operationally equivalent, denoted $[b|M]\simeq [b'|M']$, if no preparation can distinguish them, i.e.,
\begin{equation}
	\forall\, P: \quad p(b|P,M) = p(b'|P,M').
\end{equation}
In prepare-and-measure experiments, we are particularly interested in operationally equivalent procedures obtained by combining preparations $P_x$ or measurement effects $[b|M_y]$.
Specifically, one may have (hypothetical) preparations $P_\alpha=\sum_{x=1}^{n_X} \alpha_x P_x$ and $P_\beta=\sum_{x=1}^{n_X} \beta_x P_x$, where  $\{\alpha_x\}_x$ and $\{\beta_x\}_x$ are convex weights (i.e., non-negative and summing to one) and, likewise, measurement effects $[b_\alpha|M_\alpha]=\sum_{b=1}^{n_B}\sum_{y=1}^{n_Y}\alpha_{b|y}[b|M_y]$ and $[b_\beta|M_\beta]=\sum_{b=1}^{n_B}\sum_{y=1}^{n_Y}\beta_{b|y}[b|M_y]$, where $\{\alpha_{b|y}\}_{b,y}$ and $\{\beta_{b|y}\}_{b,y}$ are sets of convex weights, with $P_\alpha \simeq P_\beta$ and $[b_\alpha|M_\alpha]\simeq [b_\beta|M_\beta]$.

Such operationally equivalent procedures can naturally be grouped into equivalence classes, and it will be convenient for us to specify equivalent procedures in a slightly different, yet equivalent, way as follows.

\begin{definition}\label{defn:OE}
	\textbf{(a)} A \emph{preparation operational equivalence} is a set $\mathcal{E}_\mathcal{P}=\{(S_k,\{\xi_k(x)\}_{x\in S_k})\}_{k=1}^K$, where $\{S_k\}_k$ is a partition of $[n_X]$ into $K$ disjoint sets and, for each $k$, $\{\xi_k(x)\}_{x\in S_k}$ are convex weights (i.e., with $\xi_k(x)\ge 0$ and $\sum_{x\in S_k}\xi_k(x) = 1$).
	We say that the preparations $\{P_x\}_{x=1}^{n_X}$ satisfy $\mathcal{E}_\mathcal{P}$ if for all $k,k'\in [K]$
	\begin{equation}\label{eq:OE_P}
		\sum_{x\in S_k} \xi_k(x) P_x \simeq \sum_{x \in S_{k'}} \xi_{k'}(x)P_x.
	\end{equation}
	
	\textbf{(b)} A \emph{measurement operational equivalence} is a set $\mathcal{E}_\mathcal{M}=\{(T_\ell,\{\zeta_\ell(b,y)\}_{(b,y)\in T_\ell})\}_{\ell=1}^L$, where $\{T_\ell\}_\ell$ is a partition of $[n_B]\times [n_Y]$ into $L$ disjoint sets and, for each $\ell$, $\{\zeta_\ell(b,y)\}_{(b,y)\in T_\ell}$ are convex weights.
	We say the measurements $\{M_y\}_{y=1}^{n_Y}$ with effects $\{[b|M_y]\}_{b=1}^{n_B}$ satisfy $\mathcal{E}_\mathcal{M}$ if for all $\ell,\ell'\in [L]$
	\begin{equation}
		\sum_{(b,y)\in T_\ell} \zeta_\ell(b,y) [b|M_y] \simeq \sum_{(b,y)\in T_{\ell'}} \zeta_{\ell'}(b,y) [b|M_y].
	\end{equation}
\end{definition}

Note that any operational equivalence of the form $\sum_{x}\alpha_x P_x \simeq \sum_{x}\beta_x P_x$ or $\sum_{b,y} \alpha_{b|y}[b|M_y]\simeq \sum_{b,y}\beta_{b|y}[b|M_y]$ can be specified in this way.%
\footnote{In particular, one can obtain such a bipartition (e.g., for preparations) by taking $S_1=\{x : \alpha_x - \beta_x \ge 0\}$, $S_2=\{x : \alpha_x - \beta_x < 0\}$, $\xi_1(x) = (\alpha_x - \beta_x)/(\sum_{x\in S_1} (\alpha_x - \beta_x))$, and $\xi_2(x) = (\beta_x - \alpha_x)/(\sum_{x\in S_2} (\beta_x - \alpha_x))$.}
The formulation of Definition~\ref{defn:OE} allows us to consider natural partitions into $K\ge 2$ or $L\ge 2$ sets, which will prove useful later.
For example, if one had three operationally equivalent preparations of the form $\frac{1}{2}(P_1+P_2) \simeq \frac{1}{2}(P_3+P_4) \simeq \frac{1}{2}(P_5+P_6)$, we can express this as a single operational equivalence rather than several pairwise equivalences.

\subsection{Contextuality scenarios and noncontextuality}

With these basic notions, we can now more precisely define the kind of scenario in which we will study noncontextuality and its precise definition in such settings.
In particular, we consider prepare-and-measure scenarios of the form described above in which Alice's preparations and Bob's measurements must obey fixed sets of operational equivalences.

\begin{definition}
	A \emph{contextuality scenario} is a tuple $(n_X,n_Y,n_B,\{\mathcal{E}^{(r)}_\mathcal{P}\}_{r=1}^R,\{\mathcal{E}^{(q)}_\mathcal{M}\}_{q=1}^Q)$, where $\mathcal{E}^{(r)}_\mathcal{P}$ and $\mathcal{E}^{(q)}_\mathcal{M}$ are preparation and measurement operational equivalences, respectively.
\end{definition}

Note that the normalisation of the probability distribution $p(b|x,y)$ implies that $\sum_b [b|M_y] = \sum_b [b|M_{y'}]$ for all $y,y'$, and hence every ontological model must satisfy the corresponding operational equivalence. We will generally omit this trivial operational equivalence from the specification of a contextuality scenario.

\medskip

The notion of (operational) noncontextuality formalises the idea that operationally identical procedures must have identical representations in the underlying ontological model~\cite{Spekkens}.

\begin{definition}\label{defrealist}
An ontological model is said to be:

	\textbf{(a)} \emph{Preparation noncontextual} if it assigns the same epistemic state to operationally equivalent preparation procedures; i.e., if the preparations $P_x$ satisfy an operational equivalence $\mathcal{E}_\mathcal{P}=\{(S_k,\{\xi_k(x)\}_{x\in S_k})\}_{k=1}^K$ then, for all $k,k'\in [K]$
	\begin{equation}
		\forall \lambda: \, \sum_{x\in S_k}\xi_k(x) p(\lambda|x) = \sum_{x\in S_{k'}}\xi_{k'}(x) p(\lambda|x).
	\end{equation}
	
	\textbf{(b)} \emph{Measurement noncontextual} if it endows operationally equivalent measurement procedures with the same response function; i.e., if the measurement effects $[b|M_y]$ satisfy an operational equivalence $\mathcal{E}_\mathcal{M}=\{(T_\ell,\{\zeta_\ell(b,y)\}_{(b,y)\in T_\ell})\}_{\ell=1}^L$ then, for all $\ell,\ell'\in[L]$ 
	\begin{equation}
		\forall \lambda: \, \sum_{(b,y)\in T_\ell}\!\!\zeta_\ell(b,y) p(b|y,\lambda) =\!\! \sum_{(b,y)\in T_{\ell'}}\!\!\zeta_{\ell'}(b,y) p(b|y,\lambda).
	\end{equation}
	
	Finally, if an ontological model is both preparation and measurement noncontextual, we simply say that it is \emph{noncontextual}.
\end{definition}

The assumption of noncontextuality imposes nontrivial constraints on the probability distributions that can arise in an ontological model~\cite{Spekkens}.
\begin{definition}
	Given a contextuality scenario, the correlations $p(b|x,y)$ are said to be (preparation/measurement) noncontextual if there exists a (preparation/measurement) noncontextual ontological model satisfying the operational equivalences of the scenario and reproducing the desired correlations. If no so much model exists, we say that the correlations are \emph{(preparation/measurement) contextual}.
\end{definition}

It is known that the set of noncontextual correlations (and, likewise, the sets of preparation or measurement noncontextual correlations) forms, for a given contextuality scenario, a convex polytope delimited by \emph{noncontextuality inequalities}~\cite{NCpolytope}.

\subsection{Quantum models}

Here, we are particularly interested in what correlations can be obtained in contextuality scenarios within quantum mechanics.
In quantum theory, a preparation $P$ corresponds to a density matrix $\rho$ (i.e., satisfying $\rho \succeq 0$ and $\Tr(\rho)=1$), and two preparations $\rho$ and $\rho'$ are operationally equivalent if and only if $\rho=\rho'$.
Preparation operational equivalences thus correspond to different decompositions of the same density matrix.
Likewise, a measurement corresponds to a positive operator-valued measure (POVM) $\{E_b\}$ (defined by $E_b\succeq 0$ and $\sum_b E_b = \id$), where the $E_b$ are the measurement effects.
Measurement effects $E_b$ and $E'_{b'}$ are thus operationally equivalent if and only if $E_b = E'_{b'}$.

We can thus specify precisely what a quantum model for a contextuality scenario corresponds to.
\begin{definition}
	\label{defn:quantum_model}
	A \emph{quantum model} for a contextuality scenario $(n_X,n_Y,n_B,\{\mathcal{E}^{(r)}_\mathcal{P}\}_{r=1}^R,\{\mathcal{E}^{(q)}_\mathcal{M}\}_{q=1}^Q)$ is given by two sets of Hermitian positive semidefinite operators $\{\rho_x\}_{x=1}^{n_X}$ and $\{\{E_{b|y}\}_{b=1}^{n_B}\}_{y=1}^{n_Y}$ which satisfy
	\begin{align}
		\forall x:&\, \Tr(\rho_x)=1\label{eq:quantum_model_states}\\
		\forall y:&\, \sum_{b=1}^{n_B}E_{b|y}=\id\label{eq:quantum_model_povm}
	\end{align}
	as well as the operational equivalences
	\begin{align}
		\forall r,k:&\, \sum_{x\in S^{(r)}_k} \xi^{(r)}_k(x)\rho_x = \sigma_r \label{eq:model_OE_states}\\
		\forall q,\ell:&\, \sum_{(b,y)\in T^{(q)}_\ell} \zeta^{(q)}_\ell(b,y)E_{b|y} = \tau_q,\label{eq:model_OE_measurements}
	\end{align}
	for some operators $\sigma_r$ and $\tau_q$ independent of $k$ and $\ell$.
	
	If a quantum model consists only of pure states (i.e., if $\rho_x^2=\rho_x$ for all $x$) or projective measurements (i.e., if $E_{b|y}^2 = E_{b|y}$ and $E_{b|y}E_{b'|y}=0$ for all $b,b',y$), then we will call the model \emph{pure} or \emph{projective}, respectively.
\end{definition}

It turns out that quantum theory is conceptually different from standard realist models, in the sense that there exist quantum models for contextuality scenarios---that thus respect the specified operational equivalences---but nevertheless can give rise to contextual correlations~\cite{Spekkens}.
Quantum theory is thus said to be contextual.

Interestingly, quantum models cannot provide any advantage over noncontextual ontological models
in the absence of nontrivial preparation operational equivalences \cite{spekkens14}. In this sense, quantum theory is measurement noncontextual. Conversely, however, quantum contextuality can be witnessed in contextuality scenarios involving only operational equivalences between the preparations (along with the trivial measurement operational equivalence arising from Eq.~\eqref{eq:quantum_model_povm}, which is necessarily satisfied by any quantum model for any contextuality scenario). For this reason there has been particular interest in preparation noncontextual inequalities, although interesting contextuality scenarios involving both preparation and measurement operational equivalences have been proposed (see e.g.~\cite{NCpolytope, Mazurek2016, Kunjwal2015, ArminRoope}).

\section{A hierarchy of SDP relaxations}

In recent years, hierarchies of semidefinite programming (SDP) relaxations of the set of quantum correlations have become an invaluable tool in the study of quantum correlations~\cite{NPA1, NPA2}.
Such a hierarchy capable of bounding contextual correlations in contextuality scenarios, where operational equivalences must be taken into account, has thus far, however, proved elusive, and it is this problem we address here.

The fundamental question we are interested in is the following: given a contextuality scenario $(n_X,n_Y,n_B,\{\mathcal{E}^{(r)}_\mathcal{P}\},\{\mathcal{E}^{(q)}_\mathcal{M}\})$ and a probability distribution $p(b|x,y)$, does there exist a quantum model for the scenario reproducing the observed correlations, i.e., satisfying $p(b|x,y) = \Tr(\rho_x E_{b|y})$?

Note that, in contrast to many scenarios in quantum information, such as Bell nonlocality, it is not \emph{a priori} clear that, in the search for such a quantum model, one can restrict oneself to pure states and projective measurements despite the fact that no assumption on the Hilbert space dimension is made.
Indeed, while one can always purify a mixed state, or perform a Naimark dilation of the POVMs, such extensions may no longer satisfy the operational equivalences of the contextuality scenario.

Although SDP hierarchies have previously been formulated for prepare-and-measure scenarios \cite{NV, NV2, CharlesHierarchy, Info2}, the main challenge for contextuality scenarios is to represent the constraints arising from the operational equivalences.
Here, we adopt an approach motivated by a recent hierarchy~\cite{Info2} bounding informationally restricted correlations~\cite{Info1} and the fact that operational equivalences can be interpreted as restrictions on the information obtainable about equivalent operational procedures (see also Sec.~\ref{sec:zero-inf-games}).

\subsection{Necessary conditions for a quantum model}
\label{sec:hierarchyConds}

Similarly to other related SDP hierarchies, our approach to formulate increasingly strict necessary conditions for the existence of a quantum model is based on reformulating the problem in terms of the underlying \emph{moment matrix} of a quantum model.
To this end, let us define the set of operator variables
\begin{equation}\label{eq:operatorSet}
	J=\{\id\}\cup \{\rho_x\}_{x}\cup \{E_{b|y}\}_{b,y} \cup \{\sigma_r,\tau_\ell\}_{r,\ell},
\end{equation}
where $\sigma_r,\tau_\ell$ (with $r\in[R]$, $\ell\in[L]$) are variables corresponding to the operators defined in Eqs.~\eqref{eq:model_OE_states} and~\eqref{eq:model_OE_measurements} and will be used to enforce robustly the operational equivalences.
Consider a list $\mathcal{S}=(\mathcal{S}_1,\dots,\mathcal{S}_{|\mathcal{S}|})$ of monomials (of degree at least one) of variables in $J$. We say that $\mathcal{S}$ represents the $k$th degree of the hierarchy if it contains all monomials over $J$ of degree at most $k$.%
\footnote{In practice, however, it is often preferable to use an intermediate hierarchy level, i.e.~a monomial list that has largest degree $k$ but does not contain all terms of degree at most $k$. We will later exemplify this.} 
The choice of $\mathcal{S}$ will lead to different semidefinite relaxations, but it should at least include all elements of $J$.

Given a monomial list $\mathcal{S}$, the existence of a quantum model implies the existence of a \emph{moment matrix} $\Gamma$ whose elements, labelled by the monomials in $u,v\in\mathcal{S}$, are
\begin{equation}\label{eq:moment_matrix}
	\Gamma_{u,v} = \Tr(u^\dagger v)
\end{equation}
and satisfy a number of properties that form our necessary conditions.
Some of these constraints are common with those found in similar hierarchies (points (I)--(III) below), while others capture important aspects of quantum models for contextuality scenarios (points (IV)--(V)) and will be expressed through localising matrices~\cite{NPA3}.
We outline these constraints below.

\medskip

\noindent\textbf{(I) Hermitian positive semidefiniteness.} 
By construction the moment matrix is Hermitian and it is easily seen to be positive semidefinite~\cite{NPA1}, i.e.,
\begin{equation}\label{eq:constr_PSD}
	\Gamma = \Gamma^\dagger \succeq 0.
\end{equation}


\noindent\textbf{(II) Consistency with $p$.} 
Since the quantum model must reproduce the correlations $p(b|x,y)$, $\Gamma$ must satisfy
\begin{equation}\label{eq:constr_P}
	\forall x,y,b:\, \Gamma_{\rho_x,E_{b|y}} = p(b|x,y).
\end{equation}

\medskip

\noindent\textbf{(III) Validity of states and measurements.}
Since any quantum model must satisfy the constraints of Eqs.~\eqref{eq:quantum_model_states} and~\eqref{eq:quantum_model_povm}, $\Gamma$ must satisfy
\begin{equation}
	\forall x:\, \Gamma_{\id,\rho_x} = 1,
\end{equation}
as well as linear identities of the form
\begin{equation}\label{eq:Gamma_lin_constraints}
	\sum_{u,v}c_{u,v}\Gamma_{u,v} = 0 \quad \text{if}\quad  \sum_{u,v} c_{u,v} \Tr(u^\dagger v) = 0,
\end{equation}
where the sum is over all monomials $u,v$ in $\mathcal{S}$.
These constraints are, in particular, those satisfied by any quantum model that follow from the validity of the states and measurements making up the model and the cyclicity of the trace.

For example, Eq.~\eqref{eq:Gamma_lin_constraints} includes constraints of the form $\sum_{b}\Gamma_{E_{b|y},E_{b'|y'}}=\Gamma_{\id,E_{b'|y'}}$, as well as constraints such as $\Gamma_{E_{b|y},\rho_x E_{b'|y'}} = \Gamma_{\rho_x, E_{b'|y'}E_{b|y}}$ which follows from the fact that $\Tr(E_{b|y}\rho_x E_{b'|y'}) = \Tr(\rho_x E_{b'|y'}E_{b|y})$.
It thus includes the constraints implied by the trivial operational equivalence following from Eq.~\eqref{eq:quantum_model_povm} that are satisfied by any quantum model, thereby justifying the fact that we generally do not explicitly include this operational equivalence relation when specifying contextuality scenarios.

Note that if we were to assume the quantum model is either pure or projective (so that, respectively, either $\rho_x^2 = \rho_x$, or $E_{b|y}^2 = E_{b|y}$ and $E_{b|y}E_{b'|y}=0$), then this implies further constraints of the form~\eqref{eq:Gamma_lin_constraints}.
In particular, one can always make this assumption if there are no nontrivial operational equivalences of the corresponding type, allowing the SDP hierarchy we formulate to be simplified, but can also be considered as an additional assumption of interest (see Sec.~\ref{sec:mixedstates}).

\medskip

\noindent\textbf{(IV) Operational equivalences.}
A quantum model must satisfy the operational equivalences of Eqs.~\eqref{eq:model_OE_states} and~\eqref{eq:model_OE_measurements}. 
While this implies that the traces of each side of those equations must equal---which in turn imposes the corresponding linear identities on the moment matrix---this alone does not fully capture the constraints implied by the operational equivalences, and notably is not enough to provide a good hierarchy. To properly enforce these constraints, we draw inspiration from the hierarchy of informationally restricted quantum correlations \cite{Info2} and make use of \emph{localising matrices}. These are additional matrices of moments whose elements (or a subset thereof) are linear combinations of elements of $\Gamma$, and which themselves must be positive semidefinite~\cite{NPA3}. 

We thus define, for all $r\in[R]$, $k\in[K^{(r)}]$ and all $q\in[Q]$, $\ell\in[L^{(q)}]$, the localising matrices $\tilde{\Lambda}^{(r,k)}$ and $\hat{\Lambda}^{(q,\ell)}$ with elements
\begin{align}
	\tilde{\Lambda}^{(r,k)}_{u,v} &= \Tr\left(u^\dagger\left(\sigma_r-\sum_{x\in S_k^{(r)}}\xi_k^{(r)}(x)\rho_x\right)v\right) \\
	\hat{\Lambda}^{(q,\ell)}_{u,v} &= \Tr\left(u^\dagger\left(\tau_q-\!\!\sum_{(b,y)\in T_\ell^{(q)}}\!\!\zeta_\ell^{(q)}(b,y)E_{b|y}\right)v\right),
\end{align}
which are labelled now by monomials from a monomial list $\mathcal{L}$, in general different from $\mathcal{S}$ (and which, in principal, could differ for each localising matrix).
Ideally, $\mathcal{L}$ should be chosen so that the elements of the localising matrices are linear combinations of elements of the moment matrix $\Gamma$.

For a quantum model exactly satisfying exactly the operational equivalences $\mathcal{E}^{(r)}_\mathcal{P}$ and $\mathcal{E}^{(q)}_\mathcal{M}$, with $\sigma_r$ and $\tau_q$ defined as in Eqs.~\eqref{eq:model_OE_states} and~\eqref{eq:model_OE_measurements} one has $\tilde{\Lambda}^{(r,k)} = \hat{\Lambda}^{(q,\ell)} = 0$, $\Tr(\sigma_r)=1$ and $\Tr(\tau_q)=\sum_{(b,y)\in T_\ell^{(q)}}\zeta_\ell^{(q)}(b,y)\Tr(E_{b|y})$.
Such complicated matrix equality constraints (which one could in principle enforce without defining the localising matrices), however, tend to lead to poor results in practice due to the numerical instability of SDP solvers.
Instead, we impose the more robust constraints that $\tilde{\Lambda}^{(r,k)}, \hat{\Lambda}^{(q,\ell)}\succeq 0$ (along with the equality constraints on the traces of $\sigma_r,\tau_q$, which serve to ``normalise'' the localising matrices), which follow from the existence, for any quantum model, of Hermitian operators $\sigma_r,\tau_q$ satisfying $\sigma_r\succeq\sum_{x\in S_k^{(r)}}\xi_k^{(r)}(x)\rho_x$ and $\tau_q\succeq\sum_{(b,y)\in T_\ell^{(q)}}\zeta_\ell^{(q)}(b,y)E_{b|y}$.
We thus have, for all $r,k,q,\ell$
\begin{align}
	& \tilde{\Lambda}^{(r,k)} \succeq 0, \quad \hat{\Lambda}^{(q,\ell)} \succeq 0\\
	& \Gamma_{\id,\sigma_r} = 1, \quad \Gamma_{\id,\tau_q} =\!\!\sum_{(b,y)\in T_\ell^{(q)}}\!\!\zeta_\ell^{(q)}(b,y)\Gamma_{\id,E_{b|y}}.
\end{align}
Moreover, whenever the monomials $u$, $\sigma_r u$ and $\rho_x u$ are in $\mathcal{S}$ we have
\begin{equation}\label{eq:gamma_constr_OE_P}
	\tilde\Lambda^{(r,k)}_{u,v} = \Gamma_{u,\sigma_r v} - \sum_{x\in S^{(r)}_k} \xi^{(r)}_k \Gamma_{u,\rho_x v},
\end{equation}
and, when $u$, $\tau_q u$ and $E_{b|y} u$ are similarly in $\mathcal{S}$,
\begin{equation}\label{eq:gamma_constr_OE_M}
	\hat\Lambda^{(q,\ell)}_{u,v} = \Gamma_{u,\tau_q v} - \!\!\sum_{(b,y)\in T^{(q)}_\ell} \zeta^{(q)}_\ell \Gamma_{u,E_{b|y} v},
\end{equation}
thereby relating the localising matrices to the moment matrix $\Gamma$.

We note that the operators $\sigma_r$ and $\tau_q$, and the localising matrices expressing the deviation of their moments from those of the operational equivalences, hence play the role of slack variables to robustly enforce the operational equivalencies.
As we will see in Sec.~\ref{sec:simultingContextuality}, the formulation we adopt here will also allow a natural generalisation allowing us to study the simulation cost of preparation contextuality, where the trace of $\sigma_r$ has a natural interpretation, further motivating our choice to present the constraints in the form given here.

\medskip

\noindent\textbf{(V) Positivity of states and measurements.}
In most SDP hierarchies used in quantum information, one can assume without loss of generality that the states and measurements in question are projective (see e.g.~\cite{NPA1, NPA2}); since all projective operators are positive semidefinite, it is not necessary in such cases to consider explicitly the constraints the positive semidefiniteness of the operators in a quantum model imposes on a moment matrix.
As already mentioned, however, for contextuality scenarios this is not \emph{a priori} the case, and to capture the constraints implied by the positive semidefiniteness of states and measurements (i.e., $\rho_x, E_{b|y} \succeq 0$) we again exploit localising matrices.

Let us thus introduce the localising matrices (for all $x,y,b$) $\tilde{\Upsilon}^x$ and $\hat{\Upsilon}^{(b,y)}$ with elements
\begin{align}
	\tilde{\Upsilon}^x_{u,v} &= \Tr(u^\dagger \rho_x v)\\
	\hat{\Upsilon}^{(b,y)}_{u,v} &= \Tr(u^\dagger E_{b|y} v),
\end{align}
which are labelled by monomials from a monomial list $\mathcal{O}$, in general different from $\mathcal{S}$ (and which, as for $\mathcal{L}$, in principle could differ for each $x,y,b$).
Ideally, $\mathcal{O}$ should be chosen so that the elements of the localising matrices are also elements of the moment matrix $\Gamma$.

It is easily seen that the positive semidefiniteness of $\rho_x$ and $E_{b|y}$ implies
\begin{align}
	\forall x:&\quad \tilde{\Upsilon}^x \succeq 0\\
	\forall y,b:&\quad \hat{\Upsilon}^{(b,y)} \succeq 0,
\end{align}
which in turn (for well chosen $\mathcal{O}$) constrains $\Gamma$.
Moreover, for all $u,v$ in $\mathcal{O}$, whenever the monomials $u,\rho_x v$ or, respectively, $u,E_{b|y}v$ are in $\mathcal{S}$ we have
\begin{align}
	\tilde{\Upsilon}^x_{u,v} = \Gamma_{u,\rho_x v}, \quad \hat{\Upsilon}^{(b,y)}_{u,v} = \Gamma_{u,E_{b|y} v},
\end{align}
thereby relating the localising matrices to the main moment matrix.

\medskip

For given choices of the moment lists $\mathcal{S}$, $\mathcal{L}$ and $\mathcal{O}$, the constraints presented above thus provide necessary conditions for a given correlation to have a quantum realisation in the contextuality scenario.
Note moreover that, by standard arguments~\cite{NPA1}, one can actually assume the moment matrix (and localising matrices) are real since the above constraints only involve real coefficients.
These conditions are all semidefinite constraints, which leads us to the following proposition summarising our hierarchy of SDP relaxations.

\begin{proposition}
	\label{prop:hierarchyNC}
	Let $\mathcal{S}$, $\mathcal{L}$, $\mathcal{O}$ be fixed lists of monomials from $J$. A necessary condition for the existence of a quantum model in a given contextuality scenario reproducing the correlations $\{p(b|x,y)\}_{b,x,y}$ is the feasibility of the following SDP:
	\begin{subequations}
		\begin{align}\label{sdp}
			\textup{find} \quad &  \Gamma, \{\tilde{\Lambda}^{(r,k)}\}_{r,k}, \{\hat{\Lambda}^{(q,\ell)}\}_{q,\ell}, \{\tilde{\Upsilon}^{x}\}_{x}, \{\hat{\Upsilon}^{(b,y)}\}_{b,y} \notag \\
			\textup{s.t.} \quad & \Gamma \succeq 0,\quad \tilde{\Lambda}^{(r,k)} \succeq 0, \quad \hat{\Lambda}^{(q,\ell)} \succeq 0 \notag \\
			& \tilde{\Upsilon}^{x} \succeq 0, \quad \hat{\Upsilon}^{(b,y)} \succeq 0\\
			& \Gamma_{\rho_x,E_{b|y}}=p(b|x,y)\\
			& \Gamma_{\id,\rho_x} = 1 \\
			& \sum_{u,v}c_{u,v}\Gamma_{u,v} = 0 \quad \textup{if}\quad  \sum_{u,v} c_{u,v} \Tr(u^\dagger v) = 0 \label{eq:sdp_lin_constr}\\
			& \Gamma_{\id,\sigma_r} = 1, \quad \Gamma_{\id,\tau_q} =\!\!\sum_{(b,y)\in T_\ell^{(q)}}\!\!\zeta_\ell^{(q)}(b,y)\Gamma_{\id,E_{b|y}}\label{eq:sdp_constr_OEP}  \\
			& \tilde\Lambda^{(r,k)}_{u,v} = \Gamma_{u,\sigma_r v} - \sum_{x\in S^{(r)}_k} \xi^{(r)}_k \Gamma_{u,\rho_x v} \label{eq:sdp_constr_OEM}\\
			& \hat\Lambda^{(q,\ell)}_{u,v} = \Gamma_{u,\tau_q v} - \sum_{(b,y)\in T^{(q)}_\ell} \zeta^{(q)}_\ell \Gamma_{u,E_{b|y} v} \\
			& \tilde{\Upsilon}^x_{u,v} = \Gamma_{u,\rho_x v}, \quad \hat{\Upsilon}^{(b,y)}_{u,v} = \Gamma_{u,E_{b|y} v},\label{eq:sdp_positivity_constr}
		\end{align}
	\end{subequations}
	where the above operators are all symmetric real matrices.
\end{proposition}
By taking increasingly long monomials lists $\mathcal{S}$, $\mathcal{L}$ and $\mathcal{O}$, one thus obtains increasingly strong necessary conditions for a quantum realisation, and which can be efficiently checked by standard numerical solvers for SDPs.

While the above hierarchy applies to arbitrary contextuality scenarios, in many scenarios or situations of interest, it can be somewhat simplified.
In particular, if one wishes to determine whether a given correlation is compatible with a pure and/or projective quantum model, the extra constraints imposed on the states and measurement effects (cf.\ Definition \ref{defn:quantum_model}) correspond to further linear constraints in Eq.~\eqref{eq:sdp_lin_constr}, meaning that the corresponding localising matrices $\tilde\Upsilon^x$ and/or $\hat\Upsilon^{(b,y)}$ (and subsequent constraints in Eq.~\eqref{eq:sdp_positivity_constr}) are not required.
Similarly, if there are either no preparation or no measurement operational equivalences present in the problem (i.e., if $R=0$ or $Q=0$) then the corresponding localising matrices $\tilde\Lambda^{(r,k)}$ or $\hat\Lambda^{(q,\ell)}$ (and subsequent constraints in Eqs.~\eqref{eq:sdp_constr_OEP} and~\eqref{eq:sdp_constr_OEM}) are also not required.
The later case is particularly relevant in many (preparation) contextuality scenarios of interest, including the examples we consider in the following section.
To illustrate this, in Appendix~\ref{app:sdp_PNC} we show how the SDP simplifies for the case of preparation noncontextuality, where only nontrivial preparation operational equivalences are considered.

\medskip

Although the above hierarchy solves a feasibility problem, asking whether a distribution $p(b|x,y)$ is compatible with a quantum model for the contextuality scenario, in practice one is often interested with maximising a linear functional of the probability distribution over all possible quantum models---i.e., a \emph{noncontextuality inequality}---perhaps subject to some further constraints on the distribution.
It is easily seen that, following standard techniques, the hierarchy of necessary conditions we have presented also allows one to bound such optimisation problems by instead maximising the corresponding functional over all feasible solutions to the SDP of Proposition~\ref{prop:hierarchyNC}.

\medskip

As we will see in the following section, the hierarchy of Proposition~\ref{prop:hierarchyNC} allows us to readily obtain tight bounds on quantum contextual correlations in many scenarios of interest.
However, in some cases involving both nontrivial preparation and measurement operational equivalences and no assumptions of pure states or projective measurements, it performs relatively poorly in practice.
This appears to stem from the fact that, in such cases, the probabilities $p(b|x,y)$ do not appear on the diagonal of the moment matrix or any of the localising matrices.
In Appendix~\ref{app:sqrtHierarchy} we show how these difficulties can be overcome by presenting a modified version of our hierarchy, obtained by taking the operators $\{\sqrt{\rho_x}\}_x$ and/or $\{\sqrt{\smash[b]{E_{b|y}}}\}_{b,y}$ in the operator set $J$ (cf.\ Eq.~\eqref{eq:operatorSet}) instead of $\{\rho_x\}_x$ and $\{E_{b|y}\}_{b,y}$, an approach which we believe may be of independent technical interest.

\section{Applications of the SDP hierarchy}

We implemented a version of this hierarchy (and the variant described in Appendix~\ref{app:sqrtHierarchy}) in MATLAB, exploiting the SDP interface YALMIP~\cite{yalmip}, and our code is freely available~\cite{codeGit}.
Our implementation can handle arbitrary contextuality scenarios, restrictions to pure or projective quantum models or to classical (commuting) models, and solve either the feasibility SDP of Proposition~\ref{prop:hierarchyNC} or maximise a linear functional of the correlations $p(b|x,y)$ subject to linear constraints on the probabilities.
In solving large SDP problems that would otherwise be numerically intractable, it can make use of \mbox{RepLAB}~\cite{replab,replabGithub} (a recently developed tool for manipulating finite groups with an emphasis on SDP applications) to exploit symmetries in noncontextuality inequalities, a capability we exploit in obtaining some of the results presented below.

\subsection{Quantum violations of established preparation noncontextuality inequalities}\label{sec:qbounds}

To illustrate the usefulness of the hierarchy described in Proposition~\ref{prop:hierarchyNC}, we first exploit it to derive tight bounds on the maximal quantum violation of three preparation noncontextuality inequalities introduced in previous literature. 
In Appendix~\ref{AppHierarchyExamples} we detail the analysis of two examples based on the inequalities derived in Ref.~\cite{Ambainis} and the inequalities experimentally explored in Ref.~\cite{Hameedi}. 
Here, we focus on the noncontextuality inequalities for state discrimination presented in Ref.~\cite{Schmid}. 

To reveal a contextual advantage in state discrimination, Ref.~\cite{Schmid} considers a scenario with $x\in[4]$, $y\in[3]$ and $b\in[2]$ and attempts to discriminate the preparations $P_1$ and $P_2$, while $P_3$ and $P_4$ are symmetric extensions that ensure the operational equivalence $\frac{1}{2}P_1+\frac{1}{2}P_3\simeq\frac{1}{2}P_2+\frac{1}{2}P_4$. 
The first two measurements ($y=1,2$) correspond to distinguishing preparations $P_1$ and $P_3$, and $P_2$ and $P_4$, respectively (in the noiseless case, these should be perfectly discriminable); while the third ($y=3$) corresponds to the state discrimination task, i.e., discriminating $P_1$ and $P_2$.
There are three parameters of interest: the probability of a correct discrimination, $s$; the probability of confusing the two states, $c$; and the noise parameter, $\epsilon$. 
Under the symmetry ansatz considered, the observed statistics are thus required to satisfy
\begin{align}\nonumber
&s=p(1|1,3)=p(2|2,3)=p(2|3,3)=p(1|4,3)\\\notag
& c=p(1|2,1)=p(1|1,2)=p(2|4,1)=p(2|3,2)\\
& 1-\epsilon=p(1|2,2)=p(1|1,1)=p(2|4,2)=p(2|3,1).
\end{align}
The authors show that, for $\epsilon\leq c\leq 1-\epsilon$, the following noncontextuality inequality holds:
\begin{equation}\label{ncdiscrimination}
s\leq 1 - \frac{c-\epsilon}{2}.
\end{equation}

What is the maximal quantum advantage in the task? 
Ref.~\cite{Schmid} presented a specific family of quantum models that achieve
\begin{equation}\label{conj}
s=\frac{1}{2}\left( 1+\sqrt{1-\epsilon+2\sqrt{\epsilon(1-\epsilon)c(1-c)}+c(2\epsilon-1)}\right),
\end{equation} 
which violates the bound \eqref{ncdiscrimination}, and conjectured it to be optimal for qubit systems. 
The semidefinite programming hierarchy presented in the previous section allows us to place upper bounds on $s$ for given values of $(c,\epsilon)$ by maximising $s$ under the above constraints.
Using a moment matrix of size $42$ and localising matrices of size $7$,%
\footnote{The precise lists of moments used in this and all subsequent examples can be found along with our implementation of the SDP hierarchy, where the code generating these results is available~\cite{codeGit}. 
In all these examples we take the monomial lists $\mathcal{L}$ and $\mathcal{O}$ for the localising matrices to be the same (simply because this was sufficient to obtain the presented results), although one could indeed take these to be different if desired.} we systematically performed this maximisation with a standard numerical SDP solver~\cite{mosek}  for different values of $(c,\epsilon)$ by dividing the space of valid such parameters (i.e., satisfying $\epsilon \le c \le 1-\epsilon$) into a grid with spacing of 0.01.
We consistently obtained in every case an upper bound agreeing with the value in Eq.~\eqref{conj} to within $10^{-5}$, which is consistent with the precision of the SDP solver.
We thus find that Eq.~\eqref{conj} indeed gives the maximal quantum contextual advantage in state discrimination.

For the interested reader, in Appendix~\ref{app:tuto} we use this example to show more explicitly what form the constraints of the SDP hierarchy take and how they relate the moment matrix and localising matrices.

\subsection{Mixed states as resources for quantum contextuality}
\label{sec:mixedstates}

In many forms of nonclassicality, such as Bell nonlocality, steering and quantum dimension witnessing, the strongest quantum correlations are necessarily obtained with pure states. 
In the former two, this stems from the fact that any mixed state can be purified in a larger Hilbert space. 
In the latter, it follows from the possibility to realise a mixed state as a convex combination of pure states of the same dimension. 
Interestingly, however, it is \emph{a priori} unclear whether mixed states should play a more fundamental role in quantum contextuality: both purifications of mixed states and post-selections on pure-state components of mixed states may break the operational equivalences between preparation in contextuality scenarios. 
Here we show that this intuition turns out to be correct: preparation contextuality indeed is exceptional as mixed states are needed to obtain some contextual quantum correlations.

To prove this, we consider the noncontextuality scenario of Hameedi-Tavakoli-Marques-Bourennane (HTMB) \cite{Hameedi}. 
In this scenario, Alice receives two trits, $x\coloneqq x_1x_2\in\{0,1,2\}^2$ and Bob receives a bit $y\in[2]$ and produces a ternary outcome $b\in\{0,1,2\}$. 
There are two operational equivalences involved, corresponding to Alice sending zero information about the value of the sums $x_1+x_2$ and $x_1+2x_2$ (modulo $3$), respectively. 
Each of these corresponds to a partition of Alice's nine preparations into three sets.
Under these constraints, Alice and Bob evaluate a Random Access Code~\cite{TavakoliRAC}. 
The HTMB inequality bounds the success probability of the task in a noncontextual model \cite{Hameedi}:
\begin{equation}\label{eq:HTMB}
\mathcal{A}_\text{HTMB}\coloneqq \frac{1}{18}\sum_{x,y}p(b=x_y|x,y)\leq \frac{2}{3}.
\end{equation}

We revisit this scenario and employ our semidefinite relaxations to determine a bound on the largest value of $\mathcal{A}_\text{HTMB}$ attainable in a quantum model in which all nine preparations are pure.
As described following Proposition~\ref{prop:hierarchyNC}, this scenario can easily be considered with our hierarchy by simply including the linear constraints following from $\rho_x^2 = \rho_x$ (for all $x$) in Eq.~\eqref{eq:sdp_lin_constr} and noting that the localising matrices $\tilde\Upsilon^x$ are no longer required.
Using a moment matrix of size 2172 and localising matrices of size 187, we find that $\mathcal{A}_\text{HTMB} \lesssim 0.667$ up to solver precision.
To make such a large SDP problem numerically tractable, we used \mbox{RepLAB}~\cite{replab,replabGithub} to make the moment matrix invariant under the symmetries of the random access code, thereby significantly reducing the number of variables in the SDP problem.
This gives us strong evidence (i.e., up to numerical precision) that pure states cannot violate the HTMB inequality~\eqref{eq:HTMB}, and we conjecture this to indeed be the case exactly.%
\footnote{Note that the large size of the moment matrices meant that the solver precision we were able to obtain is somewhat reduced compared to the other examples discussed in this paper. Our numerical result agrees with the noncontextual bound of $2/3$ to within $2\times 10^{-4}$, which is within an acceptable range given the error metrics returned by the solver.} 
Importantly, however, mixed states are known to enable a violation of the inequality: six-dimensional quantum systems can achieve $\mathcal{A}_\text{HTMB}\approx 0.698$ \cite{Hameedi}.%
\footnote{We were similarly able to use our hierarchy to place an upper bound on the quantum violation of this inequality at $\mathcal{A}_\text{HTMB} \lesssim 0.704$ using a moment matrix of size 3295 and localising matrices of size 268 with the solver SCS~\cite{scs_code}. We note that obtaining this bound required using terms from the 4th level of the hierarchy. We leave it open as to what the tight quantum bound is.}
This shows that sufficiently strong contextual quantum correlations can require the use of mixed states.

\subsection{Quantum violation of contextuality inequalities involving nontrivial measurement operational equivalences}
\label{sec:measurementNC}

The examples discussed above focused on preparation contextuality scenarios, in which there are no non-trivial measurement operational equivalences. Nonetheless, quantum contextuality can also be observed in scenarios involving measurement operational equivalences (in addition to preparation operational equivalences), and we demonstrate the ability of our hierarchy to provide tight bounds in such scenarios by applying to the noncontextuality inequalities derived in Ref.~\cite{NCpolytope}.

In Ref.~\cite{NCpolytope}, the authors consider a scenario with $x\in[6]$, $y\in [3]$ and $b\in[2]$ where the preparations satisfy the operational equivalence $\frac{1}{2}(P_1 + P_2) \simeq \frac{1}{2}(P_3 + P_4) \simeq \frac{1}{2}(P_5 + P_6)$ and the measurements satisfy the operational equivalence $\frac{1}{3}\sum_{y}[1|M_y] \simeq \frac{1}{3}\sum_{y}[2|M_y]$.
The authors completely characterised the polytope of noncontextual correlations in this contextuality scenario, finding the following 6 inequivalent (under symmetries), nontrivial noncontextuality ``facet'' inequalities (where we use the notation $p_{xy}\coloneqq p(1|x,y)$):
\begin{subequations}
\begin{align}
	I_1 &= p_{11} + p_{32} + p_{53}  \le 2.5,\\
	I_2 &= p_{11} + p_{22} + p_{53}  \le 2.5,\\
	I_3 &= p_{11} - 2p_{22} + 2p_{32} - p_{41} - 2p_{51} + 2p_{53}  \le 3, \label{eq:I3}\\
	I_4 &= 2p_{11} - p_{22} + 2p_{32}  \le 3, \label{eq:I4}\\
	I_5 &= p_{11} + p_{22} + p_{32} - p_{51} + 2p_{53}  \le 4,\\
	I_6 &= p_{11} + 2p_{22} - p_{51} + 2p_{53}  \le 4.
\end{align}
\end{subequations}

While it was shown in Ref.~\cite{Mazurek2016} that a quantum model can violate the first of these inequalities and obtain the logical maximum of $I_1=3$, the degree to which the other inequalities can be violated has not, to our knowledge, previously been studied. We note that this question is also addressed in the parallel work of Ref.~\cite{Anubhav}.

In this scenario, where we have both nontrivial preparation and measurement operational equivalences, we failed to obtain nontrivial bounds on these inequalities using the basic hierarchy described by Proposition~\ref{prop:hierarchyNC}.
Instead, we employed the variant of the hierarchy described in Appendix~\ref{app:sqrtHierarchy} which uses the principal square roots of the states $\rho_x$ and/or measurements $E_{b|y}$ in the operator list, but otherwise follows the same approach.
This hierarchy, which is a strict extension of the one described by Proposition~\ref{prop:hierarchyNC}, allowed us to
place strong bounds on all the above inequalities.
Indeed, using moment matrices of size 1191 and localising matrices of size 85 (and monomials involving square roots of measurement operators, but not of states; see Appendix~\ref{app:sqrtHierarchy}), we obtained the following quantum bounds:\footnote{Due to the lack of symmetry in some of the inequalities and relatively large moment matrix size, we used the memory efficient solver SCS~\cite{scs_code} to obtain some of the results of Eq.~\eqref{eq:MNC_bounds}. This solver has the drawback of converging slower than more standard solvers~\cite{mosek}, and we thus obtain a numerical precision of the order of $10^{-3}\sim 10^{-4}$.}
\begin{equation}
	\label{eq:MNC_bounds}
	\begin{aligned}
		I_1 \lesssim 3.000,\qquad & I_2 \lesssim 2.866,\quad & I_3 \lesssim 3.500, \\
		I_4 \lesssim 3.366,\qquad & I_5 \lesssim 4.689,\quad & I_6 \lesssim 4.646.
	\end{aligned}
\end{equation}
Using a see-saw optimisation approach, for all six inequalities we were able to obtain quantum strategies saturating the bounds from the hierarchy, showing that they are in fact tight up to the precision of the SDP solver.

Interestingly, we were moreover able to show that the maximum quantum violation of the third inequality~\eqref{eq:I3} cannot be obtained with projective measurements.
Indeed, by using the hierarchy of Proposition~\ref{prop:hierarchyNC} and imposing the constraints following from the projectivity of POVM elements (and using the same monomial lists as for the above results) we were able to show that $I_3\lesssim 3.464$ for projective quantum models.
Using a see-saw optimisation, we were able to obtain projective quantum models saturating this bound to numerical precision, thereby confirming its tightness and showing that non-projective measurements, just like mixed states, are resources for quantum contextuality.

\section{Simulating preparation contextuality}
\label{sec:simultingContextuality}

Quantum correlations are famously capable of going beyond those achievable in classical theories in numerous scenarios, as highlighted by the violation of Bell inequalities and, indeed, noncontextuality inequalities.
One can likewise consider correlations that are even stronger than those observed in nature, which we call ``post-quantum'' correlations.
Interest in post-quantum theories stems from them nonetheless respecting physical principles such as no-signalling, and understanding what physical principles distinguish quantum and post-quantum correlations can lead to new insights into quantum theory itself~\cite{Popescu1994,Barrett2007,Pawlowski2009}.

An interesting strategy to study the correlations obtained by different physical theories is to ask what kind of resource, and how much of it, one should supplement a theory with to achieve stronger correlations. 
This question has been extensively studied in the context of simulating Bell correlations with classical theory and additional resources. 
Two such resources that can be used in that case are classical communication~\cite{Buhrman10,Toner} and measurement dependence~\cite{Hall11}. 
Similarly, various resources have also been investigated in Kochen-Specker contextuality experiments with the goal of simulating quantum correlations within a classical theory~\cite{Kleinmann2011,Abramsky2017,Amaral}.
To our knowledge, however, nothing is known about what resources would be necessary to simulate operationally contextual correlations, and in particular the especially relevant resource of preparation contextuality.

In this section, we begin by casting preparation contextuality scenarios as information-theoretic games, and show how these allow us to formalise a notion of simulation cost, for both classical and quantum models. 
The resource used is the preparation of states which deviate from the required operational equivalences. 
This is a natural figure of merit as the defining feature of a model for noncontextual correlations within a given theory is that the underlying ontological model obeys the specified operational equivalences;
it is thus this condition that must be violated in some way if stronger correlations are to be simulated.
We leverage our hierarchy of semidefinite relaxations to quantify both the simulation of quantum contextual correlations using classical theory, and the simulation of post-quantum correlations using quantum theory.

\subsection{Zero-information games}
\label{sec:zero-inf-games}

To show how the cost of simulating preparation contextuality can be quantified in information theoretic terms, we begin by giving an alternative interpretation for preparation contextuality scenarios (i.e., contextuality scenarios involving only nontrivial operational equivalences between sets of preparations).
In particular, we will describe how preparation contextuality scenarios can be interpreted as games in which Alice is required to hide some knowledge about her input $x$ (see, e.g., Ref.~\cite{Marvian2020}). 

Consider thus a contextuality experiment involving $R$ preparation operational equivalences.
For a given such equivalence $r\in [R]$ involving a partition into $K_r$ sets $S_k^{(r)}$, let Alice randomly choose a set $S_k^{(r)}$ (with uniform prior $p(S_k^{(r)})=1/K^{(r)}$) and a state from that set with prior $p(x|S_k^{(r)})=\xi_k^{(r)}(x)$.
How well could a receiver hope to identify which of the sets $\{S_1^{(r)},\dots,S_{K^{(r)}}^{(r)}\}$ the state they receive is sampled from?
The optimal discrimination probability in an operational theory is   
\begin{equation}\label{pg}
	G^{(r)} \coloneqq \max_{\tilde{p}(\cdot|x)}\frac{1}{K^{(r)}}\sum_{k=1}^{K^{(r)}}\sum_{x\in S_k^{(r)}}\xi_k^{(r)}(x)\,\tilde{p}(k|x)
\end{equation}
where $\tilde{p}$ is the response distribution for the discrimination. 
Using that $\sum_{k=1}^{K_r}\tilde{p}(k|x)=1$, it straightforwardly follows (see Appendix~\ref{AppInfolink}) that the discrimination probability is $G^{(r)}=\frac{1}{K^{(r)}}$ (i.e., random) if and only if the $r$th operational equivalence is satisfied. 
The discrimination probability constitutes an operational interpretation of the min-entropic accessible information about the set membership of $x$ \cite{Konig}, and is convenient to work with. More precisely, the accessible information is given by
\begin{equation}\label{info}
\mathcal{I}_r=\log_2(K^{(r)})+\log_2(G^{(r)}).
\end{equation}

Thus, we can associate the operational equivalences to an information tuple $\bar{\mathcal{I}}=\left(\mathcal{I}_1,\ldots, \mathcal{I}_R\right)$. 
A contextuality experiment is a zero-information game since $G^{(r)}=\frac{1}{K^{(r)}}$ for all $r$ is equivalent to vanishing information: $\bar{\mathcal{I}}=\bar{0}$.

\subsection{Information cost of simulating preparation contextuality}

Since a vanishing information tuple $\bar{\mathcal{I}}$ is necessary for a faithful realisation of a contextuality scenario in a given physical model, it follows that contextual correlations that cannot be explained in said model require an overhead information, i.e., an information tuple $\bar{\mathcal{I}}\neq \bar{0}$. 
In both classical (noncontextual) models and quantum theory, this means that the preparations are allowed to deviate from the operational equivalences specified by the contextuality scenario to an extent quantified by the overhead information. 
By doing so, one necessarily goes beyond a standard model for the scenario, as defined in Definition~\ref{defrealist} for classical models and Definition~\ref{defn:quantum_model} for quantum theory. 
 
For the simplest case of a single operational equivalence (i.e., $R=1$), we define the information cost, $\mathcal{Q}$, of simulating $p(b|x,y)$ in quantum theory as the smallest amount of overhead information required for quantum theory to reproduce the correlations:
\begin{align}\label{Qcost}\nonumber
& \mathcal{Q}[p]\coloneqq\min \mathcal{I} \\ \nonumber
& \text{s.t.} \quad \rho_x\succeq 0, \quad \Tr(\rho_x)=1, \quad E_{b|y}\succeq 0, \\
& \qquad \textstyle\sum_b E_{b|y}=\id,\, \text{and } \, p(b|x,y)=\Tr\left(\rho_xE_{b|y}\right).
\end{align}
However, when several operational equivalences are involved, the information is represented by a tuple $\bar{\mathcal{I}}$ and it is unclear how the information cost of simulation should be defined (note, in particular, that the operational equivalences may not be independent, so information about one may also provide information about another).
We thus focus here on the simpler case described above, and leave the more general case of $R>1$ for future research.

It is not straightforward to evaluate $\mathcal{Q}$. 
However, by modifying our semidefinite relaxations of contextual quantum correlations we can efficiently obtain lower bounds on $\mathcal{Q}$ in general scenarios. 
Indeed, note that from Eq.~\eqref{pg}, interpreted in a quantum model, it follows that if $\sigma_r$ satisfies $\sigma_r\succeq \sum_{x\in S_k^{(r)}}\xi_k^{(r)}(x)\rho_x$ for every $k\in [K^{(r)}]$ then one has $G^{(r)} \le \frac{1}{K_r}\Tr(\sigma_r)$.
Thus, rather than imposing the constraint arising from $\Tr(\sigma_r)=1$ in our hierarchy of semidefinite relaxations, we can instead minimise ($\frac{1}{K_r}$ times) the term corresponding to $\Tr(\sigma_r)$ in the moment matrix, which thus provides an upper bound on $G^{(r)}$.
Note that this provides an alternative interpretation to the constraint that $\Gamma_{\id,\sigma_r}=1$ in Eq.~\eqref{eq:sdp_constr_OEP}: it enforces the fact that Bob should have no information about which set $S_k^{(r)}$ Alice's state was chosen from.
This interpretation makes an interesting link to the recently developed approach to bounding informationally-constrained correlations~\cite{Info2}, and which indeed was the initial motivation for the approach we take in this paper.

Considering still the case of $R=1$, we thereby bound the information cost of a quantum simulation by evaluating the semidefinite relaxation as follows.

\begin{proposition}
	\label{prop:hierarchySimCost}
	For any fixed lists $\mathcal{S}$, $\mathcal{L}$, $\mathcal{O}$ of monomials from $J$, the quantum simulation cost $\mathcal{Q}[p]$ is lower bounded as
	\begin{equation}
		\log_2(K^{(1)}) + \log_2(G^*) \le \mathcal{Q}[p] ,
	\end{equation}
	where $G^*$ is obtained as
	\begin{align}
		G^* = \textup{min} \quad & \frac{\Gamma_{\id,\sigma_1}}{K^{(1)}} \\
		\textup{s.t.} \quad & \Gamma \succeq 0,\quad \tilde{\Lambda}^{(1,k)} \succeq 0, \quad \tilde{\Upsilon}^{x} \succeq 0\notag\\
		& \Gamma_{\id,\rho_x} = 1 \notag\\
		& \sum_{u,v}c_{u,v}\Gamma_{u,v} = 0 \quad \textup{if}\quad  \sum_{u,v} c_{u,v} \Tr(u^\dagger v) = 0 \notag\\
		& \tilde\Lambda^{(1,k)}_{u,v} = \Gamma_{u,\sigma_1 v} - \sum_{x\in S^{(1)}_k} \xi^{(1)}_k \Gamma_{u,\rho_x v}\notag\\
		& \tilde{\Upsilon}^x_{u,v} = \Gamma_{u,\rho_x v},\notag
	\end{align}
	where the above operators are all taken to be Hermitian.
\end{proposition}
The correctness of Proposition~\ref{prop:hierarchySimCost} follows immediately from Eq.~\eqref{info} and the fact that $G^*$ is an upper bound on $G^{(1)}$.

\medskip

Furthermore, one can similarly consider the information cost of simulation in classical models. 
In analogy with the quantum simulation cost, we define the classical simulation cost, $\mathcal{C}$, as the smallest overhead information required for a classical noncontextual model to reproduce given correlations:
\begin{align}\label{Ccost}\nonumber
& \mathcal{C}[p]\coloneqq\min \mathcal{I} \quad \text{s.t. }  \hspace{1mm}p(b|x,y)=\sum_{\lambda}p(\lambda|x)p(b|y,\lambda),\\
&\forall x:\hspace{1mm} \sum_{\lambda} p(\lambda|x)=1, \quad \forall (\lambda,y):\hspace{1mm} \sum_{b} p(b|y,\lambda)=1.
\end{align}
Naturally, in contrast to quantum simulation, every contextual distribution $p(b|x,y)$ will be associated to a non-zero classical simulation cost. 
In analogy with the quantum case, we can place lower bounds on the classical simulation cost using the SDP hierarchy we discussed and assuming that all variables commute, thereby introducing many further constraints on the SDP and providing necessary conditions for a classical model to exist for a given value of $G$. 
However, it turns out that a precise characterisation of the classical simulation cost, in terms of a linear program, is also possible by exploiting the fact that the set of classical, informationally restricted, correlations forms a convex polytope~\cite{Info1,Info2}.%
\footnote{This follows from the fact that it suffices to consider a finite alphabet size for the ontic variable $\lambda$ \cite{Info2}.} 

Finally, we make the interesting observation that the discrimination probability $G$ can be given a resource theoretic interpretation in terms of a robustness measure.
As we discuss in Appendix~\ref{AppRobustness}, this can be used to give an alternative interpretation of the simulation cost $\mathcal{I}$.

\subsection{Simulation cost in the simplest scenario}

We illustrate the above discussion of the classical and quantum simulation costs of contextuality by applying it to arguably the simplest contextuality experiment, namely parity-oblivious multiplexing (POM) \cite{POM}. 
In POM, Alice has four preparations ($x\in[4]$ written in terms of two bits $x\coloneqq x_1x_2\in[2]^2$) and Bob has two binary-outcome measurements ($y\in[2]$ and $b\in[2]$). 
The sole operational equivalence is $\frac{1}{2}P_{11}+\frac{1}{2}P_{22}\simeq\frac{1}{2}P_{12}+\frac{1}{2}P_{21}$, which corresponds to Alice's preparations carrying no information about the parity of her input $x$. 
The task is for Bob to guess the value of her $y$th input bit. 
The average success probability in a noncontextual model obeys 
\begin{equation}
\mathcal{A}_\text{POM}\coloneqq \frac{1}{8}\sum_{x,y}p(b=x_y|x,y)\leq \frac{3}{4}.
\end{equation}
In contrast, quantum models obey the tight bound $\mathcal{A}_\text{POM}\leq \frac{1}{2}\left(1+\frac{1}{\sqrt{2}}\right)$ \cite{POM}. 
However, a post-quantum probability theory can achieve the algebraically maximal success probability of $\mathcal{A}_\text{POM}=1$ \cite{Banik}.

\begin{figure}[t]
	\centering
	\includegraphics[width=\columnwidth]{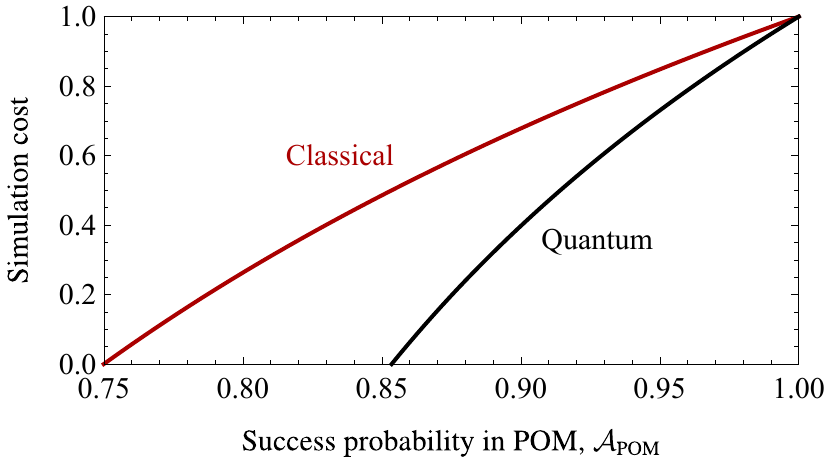}
	\caption{The information cost of simulating contextuality in parity-oblivious multiplexing using classical and quantum models.}\label{FigParity}
\end{figure}

We consider the information cost of simulating a given value of $\mathcal{A}_\text{POM}$ (i.e., the minimal information cost over all distributions compatible with that value, which can easily be evaluated by modifying the linear and semidefinite programs defined above) in both classical and quantum models. 
The results are illustrated in Fig.~\ref{FigParity}. 
The classical simulation cost is analytically given by 
\begin{equation}
	 \mathcal{C}_\text{POM} = \log_2\left(4\mathcal{A}_\text{POM}-2\right).
\end{equation}
In Appendix~\ref{AppStrats} we present an explicit simulation strategy that saturates this result, while the results of the linear program and the classical version of the hierarchy coincide with this value up to numerical precision. 

For quantum models, we have employed the described semidefinite relaxations using a moment matrix of size $547$ and localising matrices of size 89. 
The results are illustrated in Fig.~\ref{FigParity}. 
Importantly, we find that this lower bound on the quantum simulation cost is tight since we can saturate it with an explicit quantum strategy (detailed in Appendix~\ref{AppStrats}). 
The quantum simulation cost is analytically given by
\begin{equation}
\mathcal{Q}=\log_2\left(2\mathcal{A}_\text{POM}-\sqrt{2}\right)+\log_2\left(2+\sqrt{2}\right).
\end{equation}

\section{Conclusions}

In this paper we introduced a semidefinite relaxation hierarchy for bounding the set of contextual quantum correlations and demonstrated its usefulness by applying it to solve several open problems in quantum contextuality. 
This approach opens the door to the investigation of the limits of quantum contextuality in general prepare-and-measure experiments, as well as potential applications thereof. 
Moreover, it provides the building blocks with which to explore several interesting, related questions, such as whether our approach can be extended to contextuality scenarios involving more than two parties, and whether it can be adapted to bound quantum correlations in Kochen-Specker type contextuality experiments. 

By leveraging the interpretation of contextuality experiments as zero-information games, we introduced a measure of the cost of simulating preparation contextual correlations in restricted physical models, and showed how this simulation cost can be bounded in both classical and quantum models. 
This raises three fundamental questions: 
1) How can the definition of the simulation cost be extended to scenarios with multiple preparation operational equivalences which, \emph{a priori}, may not be independent?
2) How does the simulation cost of contextuality scale in prepare-and-measure scenarios with increasingly many settings? and 
3) For a given number of inputs and outputs, what is the largest simulation cost possible in order for classical correlations to reproduce quantum correlations? 
Additionally, it would be interesting to investigate how the simulation cost of operational contextuality relates to other notions of simulation, e.g., in Bell nonlocality, Kochen-Specker contextuality and communication complexity. 
In particular, can our semidefinite relaxation techniques be adapted to also bound simulation costs in such correlation experiments? 

Our work thus provides both a versatile tool for bounding quantum contextuality and a general framework for analysing the simulation of contextual correlations.

Finally, while finalising this article, we became aware of the related work of Ref.~\cite{Anubhav}. This work also addresses the problem of bounding the set of contextual quantum correlations. It uses a hierarchy of semidefinite programming relaxations that is considerably different to the one introduced here. For contextuality scenarios featuring  measurement operational equivalences, as well as general mixed states and non-projective measurements, the hierarchy of Ref.~\cite{Anubhav} appears to provide faster convergence (they recover, for example, more readily the bounds of Eq.~\eqref{eq:MNC_bounds}). In contrast, the hierarchy we introduced here appears particularly well suited to preparation contextuality scenarios, admits a generalisation to quantifying the simulation cost of contextuality, and makes an interesting conceptual connection to informationally restricted quantum correlations~\cite{Info1, Info2}.

\bigskip

\begin{acknowledgements}
The authors thank Jean-Daniel Bancal for helpful discussions on the efficient implementation of SDP hierarchies and, in particular, on the use of RepLAB to exploit symmetries in such implementations, as well as the anonymous referees for comments that helped
to significantly improve this paper. This work was supported by the Swiss National Science Foundation (Starting grant DIAQ, NCCR-SwissMAP, Early Mobility Fellowship P2GEP2 194800 and Mobility Fellowship P2GEP2 188276).
\end{acknowledgements}

\bibliography{contextuality_simulation_bib}

\appendix

\section{SDP hierarchy for preparation noncontextual correlations}
\label{app:sdp_PNC}

There has been particular interest studying noncontextual correlations in contextuality scenarios involving only nontrivial operational equivalences for preparations; i.e., preparation noncontextual correlations.
Since many of the examples we apply our SDP hierarchy to are of this form, we show here explicitly how Proposition~\ref{prop:hierarchyNC} simplifies in such scenarios.

The case of preparation noncontextuality is particularly simplified by noting that, in this particular case, we can assume the measurements to be projective.
Indeed, we can always invoke Naimark's dilation theorem to obtain projective measurements on a larger Hilbert space that give the same statistics on the states in a given quantum model.
Crucially, since there are no (nontrivial) measurement operational equivalences, these dilated projective measurements also provide a valid quantum model for the contextuality scenario in question.

\begin{proposition}
	\label{prop:hierarchyPNC}
	Let $\mathcal{S}$, $\mathcal{L}$, $\mathcal{O}$ be fixed lists of monomials from $J$. A necessary condition for the existence of a quantum model in a given preparation contextuality scenario reproducing the correlations $\{p(b|x,y)\}_{b,x,y}$ is the feasibility of the following SDP:
	\begin{subequations}
		\begin{align}
			\textup{find} \quad &  \Gamma, \{\tilde{\Lambda}^{(r,k)}\}_{r,k}, \{\tilde{\Upsilon}^{x}\}_{x} \notag \\
			\textup{s.t.} \quad & \Gamma \succeq 0,\quad \tilde{\Lambda}^{(r,k)} \succeq 0, \quad \tilde{\Upsilon}^{x} \succeq 0\\
			& \Gamma_{\rho_x,E_{b|y}}=p(b|x,y)\label{eq:sdpPNCprobs}\\
			& \Gamma_{\id,\rho_x} = 1, \quad \Gamma_{\id,\sigma_r} = 1 \\
			& \sum_{u,v}c_{u,v}\Gamma_{u,v} = 0 \quad \textup{if}\quad  \sum_{u,v} c_{u,v} \Tr(u^\dagger v) = 0 \label{eq:sdpPNC_lin_constr}\\
			& \tilde\Lambda^{(r,k)}_{u,v} = \Gamma_{u,\sigma_r v} - \sum_{x\in S^{(r)}_k} \xi^{(r)}_k \Gamma_{u,\rho_x v}\\
			& \tilde{\Upsilon}^x_{u,v} = \Gamma_{u,\rho_x v},
		\end{align}
	\end{subequations}
	where the above operators are all taken to be real symmetric matrices.
\end{proposition}

\section{Variant of SDP hierarchy using principal-square-root operators}
\label{app:sqrtHierarchy}

In the hierarchy described in Proposition~\ref{prop:hierarchyNC}, if the measurements are taken to be projective or the states pure (so they are likewise described by projectors), then all of the probabilities $p(b|x,y)$ appear on the diagonal either of one of the localising matrices $\tilde{\Upsilon}^x$ or $\hat{\Upsilon}^{(b,y)}$ or, if both these sets of operators are projective, the moment matrix $\Gamma$.
For example, in the case of projective measurements (as can always be assumed for preparation non-contextuality), one has $\tilde{\Upsilon}^x_{E_{b|y},E_{b|y}}=\Tr(E_{b|y}\rho_x E_{b|y}) = \Tr(\rho_x E_{b|y}) = p(b|x,y)$.
The positive semidefiniteness of these matrices thereby imposes strong constraints on the probability distribution, even at low levels of the hierarchy (notably, that they are non-negative, although the constraints are strictly stronger that this).

In the most general case, however, when no assumption of projective measurements or pure states can be made, the probabilities only appear on off-diagonal entries.
In practice, we found that a consequence of this was the need to go to much higher levels of the hierarchy to obtain nontrivial constraints
Indeed, for the inequalities discussed in Sec.~\ref{sec:measurementNC} we were unable to obtain useful constraints with the hierarchy of Proposition~\ref{prop:hierarchyNC}.
Here, we show how this hierarchy can be modified and generalised to overcame this shortcoming.

Our approach exploits the simple fact that, since the states $\rho_x$ and POVM elements $E_{b|y}$ are positive semidefinite, they have positive semidefinite principal square roots $\sqrt{\rho_x}$ and $\sqrt{\smash[b]{E_{b|y}}}$ such that $\sqrt{\rho_x}\sqrt{\rho_x}=\rho_x$ and $\sqrt{\smash[b]{E_{b|y}}}\sqrt{\smash[b]{E_{b|y}}}=E_{b|y}$, respectively.
Instead of taking the operator set $J$ defined in Eq.~\eqref{eq:operatorSet}, we reformulate our hierarchy using the finer-grained operator set 
\begin{equation}
	J'=\{\id \}\cup\{\sqrt{\rho_x}\}_x \cup \{\sqrt{\smash[b]{E_{b|y}}}\}_{b,y}  \cup \{\sigma_r,\tau_\ell\}_{r,\ell}.
\end{equation}
The moment matrix $\Gamma$ and localising matrices $\tilde{\Lambda}^{(r,k)},\hat{\Lambda}^{(q,\ell)}$ can be constructed in the same way as for the original hierarchy, while the localising matrices $\tilde{\Upsilon}^x$ and $\hat{\Upsilon}^{(b,y)}$ are now used to enforce the positive semidefiniteness of the principal roots, and thus have elements
\begin{align}
	\tilde{\Upsilon}^x_{u,v} &= \Tr(u^\dagger \sqrt{\rho_x} v),\\
	\hat{\Upsilon}^{(b,y)}_{u,v} &= \Tr(u^\dagger \sqrt{\smash[b]{E_{b|y}}} v).
\end{align}
While this modification may appear to change little, an immediate consequence is that the probabilities $p(b|x,y)$ now appear on the diagonal of $\Gamma$; indeed one has $\Gamma_{\sqrt{\smash[b]{E_{b|y}}}\sqrt{\rho_x},\sqrt{\smash[b]{E_{b|y}}}\sqrt{\rho_x}} = \Tr(\sqrt{\rho_x}\sqrt{\smash[b]{E_{b|y}}}\sqrt{\smash[b]{E_{b|y}}}\sqrt{\rho_x})=\Tr(E_{b|y}\rho_x)=p(b|x,y)$.

Apart from this change in operator set, the conceptual approach of the hierarchy remains unchanged.
The constraints (II)--(IV) described in Sec.~\ref{sec:hierarchyConds} are thus enforced in the same way, but now on the squares of the operators $\sqrt{\rho_x}$ and $\sqrt{\smash[b]{E_{b|y}}}$ around which the hierarchy is constructed.
For example, the constraint that, for all $x$, $\Tr(\rho_x)=1$ in any quantum model is now imposed by requiring that $\Gamma$ satisfy
\begin{equation}
	\forall x : \Gamma_{\sqrt{\rho_x},\sqrt{\rho_x}}=1.
\end{equation}

Following analogous reasoning to that of Sec.~\ref{sec:hierarchyConds}, we thus arrive at the following proposition describing the modified hierarchy.

\begin{proposition}
	\label{prop:hierarchy_sqrt_NC}
	Let $\mathcal{S}$, $\mathcal{L}$, $\mathcal{O}$ be fixed lists of monomials from $J'$. 
	A necessary condition for the existence of a quantum model in a given contextuality scenario reproducing the correlations $\{p(b|x,y)\}_{b,x,y}$ is the feasibility of the following SDP:
	\begin{subequations}
		\begin{align}\label{sdp_sqrt}
			\textup{find} \quad &  \Gamma, \{\tilde{\Lambda}^{(r,k)}\}_{r,k}, \{\hat{\Lambda}^{(q,\ell)}\}_{q,\ell}, \{\tilde{\Upsilon}^{x}\}_{x}, \{\hat{\Upsilon}^{(b,y)}\}_{b,y} \notag \\
			\textup{s.t.} \quad & \Gamma \succeq 0,\quad \tilde{\Lambda}^{(r,k)} \succeq 0, \quad \hat{\Lambda}^{(q,\ell)} \succeq 0 \notag \\
			& \tilde{\Upsilon}^{x} \succeq 0, \quad \hat{\Upsilon}^{(b,y)} \succeq 0\\
			& \Gamma_{\sqrt{\smash[b]{E_{b|y}}}\sqrt{\rho_x},\sqrt{\smash[b]{E_{b|y}}}\sqrt{\rho_x}} = p(b|x,y)\\
			& \Gamma_{\sqrt{\rho_x},\sqrt{\rho_x}} = 1 \\
			& \sum_{u,v}c_{u,v}\Gamma_{u,v} = 0 \quad \textup{if}\quad  \sum_{u,v} c_{u,v} \Tr(u^\dagger v) = 0 \\
			& \Gamma_{\id,\sigma_r} = 1\\
			& \Gamma_{\id,\tau_q}=\!\!\sum_{(b,y)\in T_\ell^{(q)}}\!\!\zeta_\ell^{(q)}(b,y)\Gamma_{\sqrt{\smash[b]{E_{b|y}}},\sqrt{\smash[b]{E_{b|y}}}}  \\
			& \tilde\Lambda^{(r,k)}_{u,v} = \Gamma_{u,\sigma_r v} - \sum_{x\in S^{(r)}_k} \xi^{(r)}_k \Gamma_{\sqrt{\rho_x}u,\sqrt{\rho_x} v} \\
			& \hat\Lambda^{(q,\ell)}_{u,v} = \Gamma_{u,\tau_q v} - \sum_{(b,y)\in T^{(q)}_\ell} \zeta^{(q)}_\ell \Gamma_{\sqrt{\smash[b]{E_{b|y}}}u,\sqrt{\smash[b]{E_{b|y}}} v} \\
			& \tilde{\Upsilon}^x_{u,v} = \Gamma_{u,\sqrt{\rho_x} v}, \quad \hat{\Upsilon}^{(b,y)}_{u,v} = \Gamma_{u,\sqrt{\smash[b]{E_{b|y}}} v},
		\end{align}
	\end{subequations}
	where the above operators are all symmetric real matrices.
\end{proposition}

Let us note firstly that Proposition~\ref{prop:hierarchy_sqrt_NC} is strictly stronger than Proposition~\ref{prop:hierarchyNC}.
Indeed, the latter can be seen as a special case of the former in which the monomial lists $\mathcal{S},\mathcal{L},\mathcal{O}$ are chosen so that the square root operators only ever appear in ``matching'' pairs.

While one may worry that one must go to higher levels of the hierarchy to obtain similarly strong constraints when employing this modified hierarchy, in practice we find that the situation is more subtle.
Even in the case where either the measurements are assumed to be projective, or the states pure, we generally found that equally tight bounds could be obtained using either hierarchy.
On the other hand, in the fully general case we found that the modified hierarchy of Proposition~\ref{prop:hierarchy_sqrt_NC} provided a clear advantage.

Finally, we note that one could likewise consider the intermediate possibility of taking the principal roots of only the states or the POVM elements in the operator set.
In this case, the probabilities instead appear on the diagonal of the localising matrices $\tilde{\Upsilon}^x$ or $\hat{\Upsilon}^{(b,y)}$.
We found that, in practice, this option generally provided the best results for moment and localising matrices of a given size.
Indeed, the results for the example of Sec.~\ref{sec:measurementNC} were obtained using the operator set
\begin{equation}
	J''=\{\id \}\cup\{\rho_x\}_x \cup \{\sqrt{\smash[b]{E_{b|y}}}\}_{b,y}  \cup \{\sigma_r,\tau_\ell\}_{r,\ell}.
\end{equation}
The implementation of our hierarchy, which is freely available~\cite{codeGit}, allows one to choose between all these different variants of the hierarchy.

We finish by noting that, to our knowledge, this approach of building an SDP hierarchy from principal square root operators is novel, at least within quantum information, and may be of independent interest in other applications.

\section{Maximal quantum violations of noncontextuality inequalities}
\label{AppHierarchyExamples}

Here we present two further case studies illustrating the practical usefulness of the hierarchy of semidefinite relaxations of the set of quantum correlations in contextuality experiments that we described in the main text.

\subsection{The inequality of Ref.~\cite{Hameedi}}

Ref.~\cite{Hameedi} experimentally implemented a test of contextuality based on the communication games introduced in Ref.~\cite{Magic7}. 
In the scenario considered there, Alice receives one of six preparations $x\coloneqq x_1 x_2$, where $x_1\in\{0,1\}$ is a bit and $x_2\in\{0,1,2\}$ a trit. 
Bob receives a binary input $y\in\{0,1\}$ and produces a ternary outcome $b\in\{0,1,2\}$. 
The authors then present the following noncontextuality inequality:
\begin{equation}
\mathcal{A}\coloneqq \frac{1}{12}\sum_{x,y,m}(-1)^m p(b=T_m| x,y)\leq \frac{1}{2},
\end{equation}
where $m=0,1$ and $T_m=x_2-(-1)^{x_1+y+m}m-x_1y \mod{3}$. This inequality is valid under the operational equivalence
\begin{equation}
\frac{1}{3}\sum_{x_2} P_{0x_2}\simeq\frac{1}{3}\sum_{x_2} P_{1x_2},
\end{equation}
i.e., when no information is relayed about the bit $x_1$. 
Notably, this noncontextuality inequality is isomorphic to the Collins-Gisin-Linden-Massar-Popescu Bell inequality \cite{CGLMP}. 
It is shown in Ref.~\cite{Hameedi} that a quantum strategy (based on qutrits) can achieve the violation $\mathcal{A}^\text{Q}=\frac{3+\sqrt{33}}{12}\approx 0.7287$, but the optimality of this violation was not proved.

Using our semidefinite relaxations we evaluated an upper bound on the largest possible value of $\mathcal{A}$ attainable in quantum theory. 
Specifically, using a moment matrix of size $386$ and localising matrices of size $49$ (with $\mathcal{L}=\mathcal{O}$) and evaluating the  corresponding semidefinite program, we obtain the value $\mathcal{A}^\text{Q}$ (up to the precision of approximately $10^{-7}$). 
Hence, up to solver precision, this shows that the quantum protocol considered in Ref.~\cite{Hameedi} is indeed optimal.

\subsection{The inequality of Ref.~\cite{Ambainis}}

Ref.~\cite{Ambainis} introduced noncontextuality inequalities based on the task of Random Access Coding. 
The authors consider a scenario in which Alice has an input $x\in[d^2]$ represented as two $d$-valued entries $x= x_1x_2\in\{0,\ldots,d-1\}^2$ while Bob receives a binary input $y\in[2]$ and produces an output $b\in\{0,\ldots, d-1\}$. 
Alice is required to communicate no information about the modular sum $x_1+x_2\mod{d}$, i.e., her preparations must respect the operational equivalences
\begin{equation}
\forall (s,s'): \quad \frac{1}{d}\sum_{x_1+x_2=s} P_x\simeq\frac{1}{d}\sum_{x_1+x_2=s'}P_x,
\end{equation}
where the addition is modulo $d$.
Ref.~\cite{Ambainis} shows that the success probability in the Random Access Code in an noncontextual model obeys
\begin{equation}
\mathcal{A}_d\coloneqq \frac{1}{2d^2}\sum_{x,y}p(b=x_y|x,y)\leq \frac{d+1}{2d}.
\end{equation}
Notably, these noncontextuality inequalities are isomorphic to known Bell inequalities for Random Access Codes \cite{Spatial}.

Let us focus on the case of $d=3$ (note that the case of $d=2$ was solved in Ref.~\cite{POM}). 
It was shown in Ref.~\cite{Ambainis} that there exists a quantum strategy (based on qutrits) which achieves the quantum violation $\mathcal{A}^\text{Q}=\frac{7}{9}$. 
However, the authors were unable to prove that a better quantum implementation cannot be found. 
Using a semidefinite relaxation corresponding to a moment matrix of size $563$ and localising matrices of size $52$ (with $\mathcal{L}=\mathcal{O}$), we evaluated an upper bound on $\mathcal{A}$ valid for general quantum models. 
Up to solver precision, we recover the result $\mathcal{A}^\text{Q}$ (it agrees up to order $10^{-8}$) thus showing that the explicit quantum strategy of Ref.~\cite{Ambainis} is optimal.

\section{Pedagogical illustration of SDP hierarchy constraints}\label{app:tuto}

To give some further understanding into the SDP hierarchy we present in Proposition~\ref{prop:hierarchyNC}, and in particular the form of the moment and localising matrices and the constraints imposed upon them, we show here somewhat more explicitly the form that they take in the example we treat in Sec.~\ref{sec:qbounds} based on state discrimination.

In this example, the only operational equivalence is the preparation operational equivalence $\frac{1}{2}P_1 + \frac{1}{2}P_3 \simeq \frac{1}{2}P_2 + \frac{1}{2}P_4$, which in the form of Definition~\ref{defn:OE} is 
\begin{equation}
	\mathcal{E}_\mathcal{P}=\big\{ \big(\{1,3\},\{\tfrac{1}{2},\tfrac{1}{2}\}\big), \big(\{2,4\},\{\tfrac{1}{2},\tfrac{1}{2}\}\big) \big\}.
\end{equation}
As a result, we assume the measurements are projective and use the simplified version of the SDP hierarchy given in Proposition~\ref{prop:hierarchyPNC}.

In order to derive the tight bounds discussed in the main text, we take the moment lists to be $\mathcal{S}=(\id, \bm{\rho}, \bm{E}, \sigma, \rho_1\bm{E}, \rho_2\bm{E}, \rho_3\bm{E}, \rho_4\bm{E}, \sigma\bm{E})$ and $\mathcal{L}=\mathcal{O} = (\id,\bm{E})$, where we use the shorthand $\bm{\rho}=(\rho_1,\rho_2,\rho_3,\rho_4)$ and $\bm{E}=(E_{1|1},E_{2|1},E_{1|2},E_{2|2},E_{1|3},E_{2|3})$, so that $\rho_x\bm{E}=(\rho_x E_{1|1},\rho_x E_{2|1},\dots)$, etc., and we denote $\sigma_1=\sigma$ since we only have one operational equivalence ($R=1$).
We thus have $|\mathcal{S}| = 42$ and $|\mathcal{L}|=|\mathcal{O}|=7$.
The moment matrix thus has the following block structure:
\begin{widetext}
	\begin{equation}
		\def\arraystretch{1.5}
		\Gamma = \left(\begin{array}{c|c|c|c|c|c|c|c|c}
		\gamma_{\id} & (1,1,1,1) & \Gamma_{\id,\bm{E}} & 1 & \bm{p}(b|1,y) & \bm{p}(b|2,y) & \bm{p}(b|3,y) & \bm{p}(b|4,y) & \Gamma_{\id,\sigma\bm{E}}\\\hline
		  & \Gamma_{\bm{\rho},\bm{\rho}} & \Gamma_{\bm{\rho},\bm{E}} & \Gamma_{\bm{\rho},\sigma} & \Gamma_{\bm{\rho},\rho_1\bm{E}} & \Gamma_{\bm{\rho},\rho_2\bm{E}} & \Gamma_{\bm{\rho},\rho_3\bm{E}} & \Gamma_{\bm{\rho},\rho_4\bm{E}} & \Gamma_{\bm{\rho},\sigma\bm{E}} \\\hline
		  && \Gamma_{\bm{E},\bm{E}} & \Gamma_{\bm{E},\sigma} & \Gamma_{\bm{E},\rho_1\bm{E}} & \Gamma_{\bm{E},\rho_2\bm{E}} & \Gamma_{\bm{E},\rho_3\bm{E}} & \Gamma_{\bm{E},\rho_4\bm{E}} & \Gamma_{\bm{E},\sigma \bm{E}} \\\hline
		  &&& \gamma_{\sigma,\sigma} & \Gamma_{\sigma,\rho_1\bm{E}} & \Gamma_{\sigma,\rho_2\bm{E}} & \Gamma_{\sigma,\rho_3\bm{E}} & \Gamma_{\sigma,\rho_4\bm{E}} & \Gamma_{\sigma,\sigma\bm{E}} \\\hline
		  &&&& \Gamma_{\rho_1\bm{E},\rho_1\bm{E}} & \Gamma_{\rho_1\bm{E},\rho_2\bm{E}} & \Gamma_{\rho_1\bm{E},\rho_3\bm{E}} & \Gamma_{\rho_1\bm{E},\rho_4\bm{E}} & \Gamma_{\rho_1\bm{E},\sigma\bm{E}} \\\hline
		  &&&&& \Gamma_{\rho_2\bm{E},\rho_2\bm{E}} & \Gamma_{\rho_2\bm{E},\rho_3\bm{E}} & \Gamma_{\rho_2\bm{E},\rho_4\bm{E}} & \Gamma_{\rho_2\bm{E},\sigma\bm{E}} \\\hline
		  &&&&&& \Gamma_{\rho_3\bm{E},\rho_3\bm{E}} & \Gamma_{\rho_3\bm{E},\rho_4\bm{E}} & \Gamma_{\rho_3\bm{E},\sigma\bm{E}} \\\hline
		  &&&&&&& \Gamma_{\rho_4\bm{E},\rho_4\bm{E}} & \Gamma_{\rho_4\bm{E},\sigma\bm{E}} \\\hline
		  &&&&&&&& \Gamma_{\sigma\bm{E},\sigma\bm{E}}
		\end{array}\right),
	\end{equation}
\end{widetext}
where the blocks correspond to the block specification of $\mathcal{S}$ given above, and we have given only the upper triangle since the matrix is symmetric.
The vectors $\bm{p}(b|x,y)$ are to be understood as $\bm{p}(b|x,y)=(p(1|x,1),p(2|x,1),p(1|x,2),p(2|x,2),p(1|x,3),p(2|x,3))$.
Let us note immediately that, by Eq.~\eqref{eq:sdpPNCprobs}, one has
\begin{equation}
	\Gamma_{\rho,E} = \begin{pmatrix}
		\bm{p}(1|1,y) \\ 
		\bm{p}(2|1,y)\\
		\bm{p}(3|1,y)\\
		\bm{p}(4|1,y)
	\end{pmatrix}.
\end{equation}

The localising matrices $\tilde{\Upsilon}^x$ and $\tilde{\Lambda}^{(k)}$ can readily be identified as 
\begin{equation}
	\tilde{\Upsilon}^x = \left(
	\begin{array}{c|c}
		\gamma_{\id} & \bm{p}(b|x,y) \\\hline
		& \Gamma_{\bm{E},\rho_x\bm{E}}
	\end{array}
	\right)
\end{equation}
and
\begin{equation}
	\tilde{\Lambda}^{(k)} = \left(
	\begin{array}{c|c}
		\gamma_{\id} & \Gamma_{\sigma\bm{E}} \\\hline
		& \Gamma_{\bm{E},\sigma\bm{E}}
	\end{array}
	\right)
	- \frac{1}{2}\sum_{x\in S_k}
	\tilde{\Upsilon}^x,
\end{equation}
where $S_1=\{1,3\}$ and $S_2=\{2,4\}$.

The remaining constraints of interest are those referred to in Eq.~\eqref{eq:sdpPNC_lin_constr}.
To illustrate these, let us expand on the form of some of the blocks in $\Gamma$.
From the completeness relation $\sum_bE_{b|y}=\id$ we can write $\Gamma_{\bm{\rho},\sigma\bm{E}}$ as
\begin{widetext}
\begin{equation}
	\Gamma_{\bm{\rho},\sigma\bm{E}} =
	\begin{pmatrix}
		\gamma_{\rho_1\sigma E_{1|1}}\  & \gamma_{\rho_1\sigma} - \gamma_{\rho_1\sigma E_{1|1}}\  & \gamma_{\rho_1\sigma E_{1|2}}\  & \gamma_{\rho_1\sigma} - \gamma_{\rho_1\sigma E_{1|2}}\  & \gamma_{\rho_1\sigma E_{1|3}}\  & \gamma_{\rho_1\sigma} - \gamma_{\rho_1\sigma E_{1|3}} \\
		\gamma_{\rho_2\sigma E_{1|1}}\  & \gamma_{\rho_2\sigma} - \gamma_{\rho_2\sigma E_{1|1}}\  & \gamma_{\rho_2\sigma E_{1|2}}\  & \gamma_{\rho_2\sigma} - \gamma_{\rho_2\sigma E_{1|2}}\  & \gamma_{\rho_2\sigma E_{1|3}}\  & \gamma_{\rho_2\sigma} - \gamma_{\rho_2\sigma E_{1|3}} \\
		\gamma_{\rho_3\sigma E_{1|1}}\  & \gamma_{\rho_3\sigma} - \gamma_{\rho_3\sigma E_{1|1}}\  & \gamma_{\rho_3\sigma E_{1|2}}\  & \gamma_{\rho_3\sigma} - \gamma_{\rho_3\sigma E_{1|2}}\  & \gamma_{\rho_3\sigma E_{1|3}}\  & \gamma_{\rho_3\sigma} - \gamma_{\rho_3\sigma E_{1|3}} \\
		\gamma_{\rho_4\sigma E_{1|1}}\  & \gamma_{\rho_4\sigma} - \gamma_{\rho_4\sigma E_{1|1}}\  & \gamma_{\rho_4\sigma E_{1|2}}\  & \gamma_{\rho_4\sigma} - \gamma_{\rho_4\sigma E_{1|2}} \ & \gamma_{\rho_4\sigma E_{1|3}}\  & \gamma_{\rho_4\sigma} - \gamma_{\rho_4\sigma E_{1|3}} 
	\end{pmatrix},
\end{equation}
where $\gamma_{\rho_x\sigma}$ are the elements of $\Gamma_{\bm{\rho},\sigma}$.
By the cyclicity of the trace and the projectivity of the measurement (i.e., $E_{b|y}E_{b'|y}=\delta_{b,b'}E_{b|y}$), the elements of $\Gamma_{\bm{\rho},\sigma\bm{E}}$ are then related to the elements of $\Gamma_{\rho_x \bm{E},\sigma{\color{red}\bm{E}}}$ as (recalling that $\Gamma_{u,v} = \Tr(u^\dagger v)$, so the elements of the monomial $u$ are reversed)
\begin{align*}
	\Gamma_{\rho_x \bm{E},\sigma\bm{E}} &= 
	\begin{pmatrix}
		\gamma_{\rho_x\sigma E_{1|1}} & 0 & \gamma_{\rho_x\sigma E_{1|2}E_{1|1}} & \gamma_{\rho_x\sigma E_{2|2}E_{1|1}} & \gamma_{\rho_x\sigma E_{1|3}E_{1|1}} & \gamma_{\rho_x\sigma E_{2|3}E_{1|1}}\\
		0 & \gamma_{\rho_x\sigma E_{2|1}} & \gamma_{\rho_x\sigma E_{1|2}E_{2|1}} & \gamma_{\rho_x\sigma E_{2|2}E_{2|1}} & \gamma_{\rho_x\sigma E_{1|3}E_{2|1}} & \gamma_{\rho_x\sigma E_{2|3}E_{2|1}}\\
		\gamma_{\rho_x\sigma E_{1|1}E_{1|2}} & \gamma_{\rho_x\sigma E_{2|1}E_{1|2}} & \gamma_{\rho_x\sigma E_{1|2}} & 0 & \gamma_{\rho_x\sigma E_{1|3}E_{1|2}} & \gamma_{\rho_x\sigma E_{2|3}E_{1|2}}\\
		\gamma_{\rho_x\sigma E_{1|1}E_{2|2}} & \gamma_{\rho_x\sigma E_{2|1}E_{2|2}} & 0 & \gamma_{\rho_x\sigma E_{2|2}} & \gamma_{\rho_x\sigma E_{1|3}E_{2|2}} & \gamma_{\rho_x\sigma E_{2|3}E_{2|2}}\\
		\gamma_{\rho_x\sigma E_{1|1}E_{1|3}} & \gamma_{\rho_x\sigma E_{2|1}E_{1|3}} & \gamma_{\rho_x\sigma E_{1|2}E_{1|3}} & \gamma_{\rho_x\sigma E_{2|2}E_{1|3}} & \gamma_{\rho_x\sigma E_{1|3}} & 0 \\
		\gamma_{\rho_x\sigma E_{1|1}E_{2|3}} & \gamma_{\rho_x\sigma E_{2|1}E_{2|3}} & \gamma_{\rho_x\sigma E_{1|2}E_{2|3}} & \gamma_{\rho_x\sigma E_{2|2}E_{2|3}} & 0 & \gamma_{\rho_x\sigma E_{2|3}},
	\end{pmatrix}
\end{align*}
\end{widetext}
where, for the sake of legibility, we have not yet applied the completeness relations.
These, e.g., further impose that $\gamma_{\rho_x\sigma E_{2|2}E_{1|1}} = \gamma_{\rho_x\sigma E_{1|1}} - \gamma_{\rho_x\sigma E_{1|2}E_{1|1}}$, $\gamma_{\rho_x\sigma E_{2|2}E_{2|1}} = \gamma_{\rho_x\sigma} - \gamma_{\rho_x\sigma E_{1|2}} - \gamma_{\rho_x\sigma E_{1|1}} + \gamma_{\rho_x\sigma E_{1|2}E_{1|1}}$, etc.

The other blocks of $\Gamma$ can be reduced and related in similar ways by applying similar simplifications.

In practice, our code (which is freely accessible~\cite{codeGit}), works by applying reductions to every element of the moment matrix to reduce it to a canonical form, before identifying the unique elements.
The completeness relations can then be applied to further reduce the number of variables in the optimisation problem.
We note, however, that when projective measurements are considered it is generally not actually necessary to apply the constraints arising from the completeness relation.
Although one obtains a potentially weaker set of necessary conditions, in practice we rarely see any difference in the power of the hierarchy under this relaxation.

\section{Contextuality experiments as zero-information games}\label{AppInfolink}

Here we show that, for a given operational equivalence (i.e., a fixed $r\in [R]$), a uniform discrimination probability $G=1/K$ (i.e., a vanishing information $\mathcal{I}=0$) is equivalent to the corresponding operational equivalence
\begin{equation}\label{eq:OE_P2}
\sum_{x\in S_k(x)} \xi_k(x) P_x = \sum_{x \in S_{k'}(x)} \xi_{k'}(x)P_x
\end{equation}
being satisfied. 
To this end, use that $\sum_{k=1}^K\tilde{p}(k|x)=1$ to write the discrimination probability on the form
\begin{align}
G=\frac{1}{K}+
\max_{\tilde{p}(\cdot|x)}\frac{1}{K}\sum_{k=1}^{K-1}\Bigg[&\sum_{x\in S_k(x)} \xi_k(x) \tilde{p}(k|x) \notag\\
& -\!\sum_{x\in S_K(x)} \xi_K(x) \tilde{p}(k|x)\Bigg].
\end{align}
It then follows from the convex linearity of $\tilde{p}$ in $x$ (cf.\ Footnote~\ref{fn:CL}, noting that $\tilde{p}$ must by definition arise from an ontological model) that the operational equivalences \eqref{eq:OE_P2} imply that the bracket in the above expression vanishes, thus leading to $G=1/K$. Conversely, the condition $G=1/K$ is equivalent to 
\begin{equation}
0=\max_{\tilde{p}(\cdot|x)}\sum_{k=1}^{K-1}\left[\sum_{x\in S_k(x)} \xi_k(x) \tilde{p}(k|x)-\!\!\sum_{x\in S_K(x)}\!\! \xi_K(x) \tilde{p}(k|x)\right].
\end{equation}
If the bracket on the right-hand-side does not vanish we can always find a $\tilde{p}(\cdot|x)$ such the argument of the maximisation becomes positive. Thus, the operational equivalences are implied.

\section{Simulation cost from robustness of operational inequivalence}\label{AppRobustness}

Let us first show how the discrimination probability $G$, as defined in Eq.~\eqref{pg}, can be related to a robustness measure within a resource-theoretic framework (see Ref.~\cite{regula17} for an overview of robustness measures in such frameworks).
To start with, let us write $G$, for a quantum model, as
\begin{align}\label{Sucprob}
G&=\max_M \frac{1}{K}\sum_{k=1}^{K}\sum_{x\in S_k}\xi_k(x)\Tr\left[\rho_x E_k\right]\nonumber\\
&=\max_M \frac{1}{K}\sum_{k=1}^K \Tr\left[\hat\sigma_{k} E_k\right],
\end{align}
where $\hat\sigma_{k}\coloneqq \sum_{x\in S_k}\xi_k(x)\rho_x$ and the maximisation is taken over POVMs $\{E_k\}_k$. 
Note that this optimisation  can equally well be written as an optimisation over sets of operators $\{E_k\}_k$ for which $E_k\succeq 0$ for all $k$ and $\sum_k\Tr[\sigma E_k]\leq 1$ for all quantum states $\sigma$. 
In this way, the optimisation variables $\{E_k\}_k$ can be interpreted as witnesses for the (non-membership in the) set $F$ consisting of tuples of states of the form $\{\hat\sigma_k\}_k=\{\sigma,\dots,\sigma\}$, where $\sigma$ is some quantum state. 
This ``free set'' $F$ can be interpreted as the set of coarse grained states $\sum_x\xi_k^{(r)}(x)\rho_x$ arising from operationally equivalent preparations.
Indeed, any such preparations give, by definition, $\{\hat\sigma_k\}_k\in F$, while any $\{\sigma,\dots,\sigma\}\in F$ can be obtained from such a set of preparations by taking $\rho_x=\sigma$ for all $x$.

Given a witness for a set, one can look to give an interpretation to its violation.
Such an interpretation depends highly on the form of the involved optimisation problem. 
In our case, our optimisation problem is formulated in such a way that it corresponds to a commonly used resource quantifier in quantum resource theories. 
Namely, the problem in Eq.~\eqref{Sucprob} is, up to scaling and shifting, the SDP dual of the generalised robustness measure $R_F$ of a state tuple $\mathcal T=\{\hat\sigma_k\}_k$ with respect to the set $F$. 
The generalised robustness is defined as
\begin{align}
R_F(\mathcal T)\coloneqq\min\Big\{t\geq0\mid\frac{\mathcal T+t\tilde{\mathcal T}}{1+t}\in F\Big\},
\end{align}
where the optimisation is over all tuples of states $\tilde{\mathcal T}$ that are of the same size as $\mathcal T$. 
$R_F$ is thus a measure of how much the operational equivalences between the states $\{\rho_x\}_x$ are violated (or, more precisely, how far their coarse-grainings are from being equal), and we call it the (generalised) robustness of operational inequivalence.
The dual of $R_F(\mathcal T)$ is straightforward to obtain (see, e.g., Ref.~\cite{Takagi19,Uola19}), leading to the relation
\begin{align}
    1+R_F(\mathcal T)= K\, G.
\end{align}
Hence, for a given set of states $\{\rho_x\}_x$, the accessible information $\mathcal{I}$ can be related to the robustness of the corresponding tuple $\{\hat\sigma_k\}_k$ with respect to those tuples that contain no information about the index $k$, i.e.
\begin{align}
\mathcal{I}=\log(1+R_F(\mathcal T)).
\end{align}
In this way, the quantum simulation cost corresponds to the minimum robustness $R_F(\mathcal T)$ taken over all tuples $\mathcal T$ arising from states $\{\rho_x\}_x$ compatible with the observed statistics $p(b|x,y)$.

We note that, in recent years, several links between robustness measures and advantages in discrimination tasks~\cite{Takagi19,Uola19} and more general quantum games~\cite{Uola20} have been uncovered.
Free sets of the form of $F$ have not previously been studied, and it remains an interesting open question to study how the robustness of operation inequivalence relates to advantages in such operational tasks, albeit one beyond the scope of the present paper.

\section{Simulation strategies}\label{AppStrats}

Here we first present an optimal strategies for both the classical and quantum simulation cost of contextual correlations in parity-oblivious multiplexing (POM).

\subsection{Classical model}

We first give an optimal strategy for the classical simulation cost of parity-oblivious multiplexing.
A score of $\mathcal{A}_\text{POM}=3/4$ can be obtained by the trivial classical strategy of sending only $x_1$, which reveals no information about the parity of $x_1\oplus x_2$.
A score $\mathcal{A}_\text{POM} > 3/4$ can be obtained by mixing this strategy (which has a simulation cost of $0$) with another trivial strategy that has a simulation cost of $1$ (e.g., by classically sending both $x_1$ and $x_2$).
More precisely, a score of
\begin{equation}\label{eq:APOM_classical_lin}
	\mathcal{A}_\text{POM} = 1-q/4,
\end{equation}
for $q\in[0,1]$, can be obtained by using these two strategies with probabilities $q$ and $1-q$, respectively.
This trivially gives $G=1-\frac{q}{2}$ and thus $\mathcal{I}=\log_2(2-q)$, providing an upper bound on $\mathcal{C}$.

Using the hierarchy of semidefinite relaxations with commuting operators as described in the main text, we tested 100 values of $\mathcal{A}_\text{POM}$ as given in Eq.~\eqref{eq:APOM_classical_lin} (with $q\in[0,1]$) and found that, in each case, the lower bound on $\mathcal{C}$ matched the above upper bound (up to the numerical precision of the solver).
We thus find, as claimed in the main text, that
\begin{equation}
\mathcal{C}=\log_2\left(4\mathcal{A}_\text{POM} - 2\right).
\end{equation}

\subsection{Quantum model}

Here we describe an optimal quantum strategy for simulating post-quantum correlations in parity-oblivious multiplexing.
Recall that the optimal quantum strategy in parity-oblivious multiplexing gives $\mathcal{A}_\text{POM}=\frac{1}{2}(1+\frac{1}{\sqrt{2}})$.
This value can be obtained by taking the preparations 
\begin{equation}
	\rho_{x_1 x_2}=\frac{1}{2}\left(\id + \frac{(-1)^{x_1}\sigma_x + (-1)^{x_2}\sigma_y}{\sqrt{2}}\right)
\end{equation}
and performing projective measurements in the $x$- and $y$-bases if $y=1$ or $y=2$, respectively~\cite{POM}.

A score $\mathcal{A}_\text{POM} > \frac{1}{2}(1+\frac{1}{\sqrt{2}})$ can be obtained by mixing this optimal strategy (which has a simulation cost of $0$ as it obeys the operational equivalences), and a trivial strategy with a simulation cost of $1$ (e.g., by classically sending both $x_1$ and $x_2$).
More precisely, a score of
\begin{equation}\label{eq:APOM_lin}
	\mathcal{A}_\text{POM} = 1-\frac{q}{2}\left(1-\frac{1}{\sqrt{2}}\right),
\end{equation}
for $q\in[0,1]$, can be obtained by using these two strategies with probabilities $q$ and $1-q$, respectively.
This trivially gives $G=1-\frac{q}{2}$ and thus $\mathcal{I}=\log_2(2-q)$, providing an upper bound on $\mathcal{Q}$.

Using the hierarchy of semidefinite relaxations as described in the main text, we tested 100 values of $\mathcal{A}_\text{POM}$ as given in Eq.~\eqref{eq:APOM_lin} (with $q\in[0,1]$) and found that, in each case, the lower bound on $\mathcal{Q}$ matched the above upper bound (up to the numerical precision of the solver).
We thus find, as claimed in the main text, that
\begin{equation}
\mathcal{Q}=\log_2\left(2\mathcal{A}_\text{POM}-\sqrt{2}\right)+\log_2\left(2+\sqrt{2}\right).
\end{equation}

\end{document}